\documentclass[aps, preprint, eqsecnum, showpacs, epsfig]{revtex4}
\usepackage{array,longtable}
\usepackage{graphics}
\usepackage{graphicx}
\usepackage{epsfig}
\usepackage{amsmath}
\newcommand{\be}{\begin{equation}}
\newcommand{\ee}{\end{equation}}
\newcommand{\bse}{\begin{subequations}}
\newcommand{\ese}{\end{subequations}}
\def \vr {{\roarrow r}}
\def \vG {{\roarrow G}}
\def \vR {{\roarrow R}}
\def \vq {{\roarrow q}}
\def \rhor{{$ \rho(\roarrow r) $ }}
\def \rhol{$ \rho_{l} $}
\def \rhos{$ \rho_{0} $}
\def \op{$ \rho_{G} $}
\def \gaml{$ \gamma_{l} $}
\def \gams{$ \gamma_{s} $}
\def \dga{$ \Delta \gamma $}
\def \degree{^{\circ}}

\begin{document}
\title{Correlation functions in liquids and crystals : Free energy functional and liquid - crystal transition}
\author{Atul S. Bharadwaj, Swarn L. Singh$^{*}$ and Yashwant Singh}
\author{}
\affiliation{Department of Physics, Banaras Hindu University, 
Varanasi-221 005, India.}
\date{\today}

\begin{abstract}
  A free energy functional for a crystal that contains both the symmetry conserved and 
 symmetry broken parts of the direct pair correlation function 
 has been used to investigate the crystallization
 of fluids in three-dimensions. The symmetry broken part of the direct pair correlation function
 has been calculated using a series in ascending powers of the order parameters and which contains
 three- and higher-bodies direct correlation functions of the isotropic phase. It is shown that 
 a very accurate description of freezing transitions for a wide class of potentials is found by
  considering the first two terms of this series. The results found for freezing parameters
  including structure of the frozen phase for fluids interacting via the
 inverse power potential $ u(r) = \epsilon \left(\sigma/r\right)^{n}$ for n ranging from $4$
 to $\infty$ are in very good agreement with simulation results. It is found that for $n > 6.5$
 the fluid freezes into a face centred cubic (fcc) structure while for $n \leq 6$ the body centred
 cubic (bcc) structure is preferred. The fluid-bcc-fcc triple point is found to be at $1/n = 0.158$
 which is in good agreement with simulation result.
\end{abstract}

\footnotetext[1]{$^{*}$~Present address-Max Planck Institute for Intelligent Systems,
Heisenbergstr. 3, 70569 Stuttgart, Germany,
and Institut f\"{u}r Theoretische Physik IV, Universit\"{a}t  Stuttgart, 
Pfaffenwaldring 57, 70569 Stuttgart, Germany.}


\maketitle

\section{ Introduction}

 Freezing of a fluid into a crystalline solid is a particular, but an important example of a first-order phase
 transition in which the continuous symmetry of the fluid is broken into one of the Bravais lattices. The 
 transition in three-dimensions is marked by large discontinuities in entropy, density and order parameters; 
 the order parameters being proportional to the lattice components of one particle density distribution \rhor 
 (see Eq.(2.3)). Efforts have been made for over six decades \cite{1,2} to find a first principle 
 theory which can answer 
 questions such as, at what density, pressure and temperature does a particular liquid freeze ? 
 What is the change in entropy and the change in density upon freezing ? 
 Which of the Bravais lattices emerges at the freezing point for a given system and 
 what are values of the order parameters?

 A crystal is a system of extreme inhomogeneities where value of \rhor shows several orders of magnitude 
 difference between its values on the lattice sites and in the interstitial regions. The density functional 
 formalism of classical statistical mechanics has been employed to develop theories for freezing 
 transitions \cite{2,3}. 
 This kind of approach was initiated in 1979 by Ramakrishnan and Yussouff (RY) \cite{4} which was latter reformulated 
 by Haymet and Oxtoby \cite{5}. The central quantity in this formalism is the reduced Helmholtz free energy of both 
 the crystal, $ A[\rho] $, and the liquid , $ A(\rho_{l})$ \cite{2}. For crystals, $ A[\rho] $ is a unique functional 
 of \rhor whereas for liquids, $ A(\rho_{l})$ is simply a function of liquid density \rhol\ which is a constant, 
 independent of position.

 The density functional formalism is used to write an expression for $ A[\rho] $ (or for the grand thermodynamic 
 potential) in terms of \rhor and the direct pair correlation function (DPCF). Minimization of this expression 
 with respect to \rhor leads to an expression that relates \rhor to the DPCF. The DPCF that appears in these
  equations corresponds to the crystal and is functional of \rhor and therefore depends on values of
   the order parameters.
  In the RY theory the functional dependence of  the DPCF on \rhor  was neglected and was replaced by that of 
  the coexisting liquid of density \rhol. Attempts to improve the RY theory by incorporating a term 
  involving three-body direct correlation function of the coexisting liquid in the expression of 
  $ A[\rho] $ have failed \cite{6,7}. The efforts made by Tarazona \cite{8}, Curtin, Ashcraft and
  Denton \cite{9,10} and others \cite{11,12} in the direction of developing a theory using what
  is referred to as the \textit{weighted density approximation} have also met with limited
  success only.

 The reason, as has been pointed out recently \cite{13,14}, is that at the fluid - solid transition the isotropy and the 
 homogeneity of space is spontaneously broken and a qualitatively new contribution to the correlation in 
 distribution of particles emerges. This fact has been used to write the DPCF of the frozen phase as a sum of 
 two terms; one that preserves the continuous symmetry of the liquid and the other that breaks it and
  vanishes in the liquid. An exact expression for the free energy functional was found by performing double 
  functional integration in density space of a relation that relates the second functional
  derivative of $ A[\rho] $ with respect to \rhor\ to the DPCF (see Eq.(\ref{2.7})). 
  This  expression of free energy functional contains both the symmetry conserved and the symmetry broken parts 
  of the  DPCF.

 The values of the DPCF as well as of the total pair correlation function (described in Sec II) in a classical 
 system can be found from solution of integral equation, the Ornstein - Zernike (OZ) equation, and a closure 
 relation that relates correlation functions to pair potential \cite{15}. The integral equation theory has been 
 quite successful in getting values of pair correlation functions of uniform liquids \cite{15}, but its 
 application to find pair correlation functions of symmetry broken phases has so far been limited. Recently 
 Mishra and Singh \cite{16} have used the OZ equation and the Percus - Yevick (PY) closure relation to obtain both the 
 symmetry conserved and symmetry broken parts of pair correlation functions in a nematic phase. In the nematic 
 phase the orientational symmetry is broken but the translational symmetry of the 
 liquid phase remains intact whereas in a crystal both the orientational and the translational 
 symmetries of the liquid phase are broken. Since, closure relations are derived assuming
  translational invariance \cite{15}, they are valid in normal liquids as well as in nematics 
  but may not in crystals. In view of this, Singh and Singh \cite{13} suggested a method in which the 
 symmetry broken part of the DPCF is expanded in ascending powers of order parameters. This series contains 
 three- and higher - bodies direct correlation functions of the isotropic phase. The first term of this series 
 was evaluated and used in investigating the freezing transitions in two- and three-dimensions of fluids 
 interacting via inverse power potentials \cite{13,17} and freezing of hard spheres into crystalline and 
 glassy phases \cite{14}. It has been found that contribution made by the symmetry broken part to the 
 grand thermodynamic potential at the freezing point increases with softness of the potential \cite{13,17}. 
 This suggests that for long - ranged potentials the higher order terms of the series may not be negligible and 
 need to be considered.

 In this paper we calculate first and second terms of the series (see Eq.(2.29)) which involve three and 
 four-bodies direct correlation functions of the isotropic phase. We calculate the four-body direct 
 correlation function by extending the method developed to 
 calculate the three-body direct correlation function. The values found for the DPCF are used in the free-energy 
 functional and the crystallization of fluids is investigated. We show that all questions posed at the beginning 
 of this section are correctly answered for a wide class of potentials.

 The paper is organised as follows: In Sec. II we describe correlation functions in liquids and in crystals and 
 calculate them. The symmetry broken part of the DPCF is evaluated  using first two terms of a series in 
 ascending powers of order parameters. 
 These results are used in the free-energy functional in Sec. III to calculate 
 the contributions made by different parts of the DPCF to the grand thermodynamic potential at the freezing 
 point. In Sec. IV we calculate these terms and locate the freezing points for fluids interacting via the 
 inverse power potentials and compare our results with those found from computer simulations and from 
 approximate free energy functionals. The paper 
 ends with a brief summary and perspectives given in Sec. V.

\section{  Correlation Functions}

The equilibrium one particle distribution \rhor defined as
\be
\rho(\vr) = \left\langle \sum_{l} \delta \left( \vr - \vr_{l} \right)\right\rangle , 
\ee
 where $ \vr_{l} $ is position vector of the $ l^{th} $ particle and the angular bracket, $ \left\langle
 ....\right\rangle  $, represents the ensemble average, is a constant, independent of position for a normal 
 liquid but contains most of the structural informations of a crystal. For a crystalline solid there exists a 
 discrete set of vectors $ \vR_{i} $ such that,

\be
\rho(\vr) = \rho(\vr + \vR_{i}),  \qquad \qquad
\ee 
\vspace{-1.85cm}
\begin{center}
  \hspace{2.0in} for all $ \vR_{i}$.
\end{center}           

 This set of vectors which appears at the freezing point due to spontaneous breaking of continuous symmetry of 
 a liquid, necessarily forms a Bravais lattice. The \rhor in a crystal can be written as a sum of two terms:
 
\bse
\begin{align}
\rho(\vr) = \rho_{0} + \rho^{(b)} (\vr) \label{2.3a} 
\end{align}
where 
\begin{align}
\rho^{(b)}(\vr) = \sum_{G} \rho_{G} e^{i\vG.\vr}. \label{2.3b} 
\end{align}
\ese

 Here \rhos\ is the average density of the crystal and \op\ are the order parameters 
 (amplitude of density waves of wavelength $ 2\pi/\vert \vG \vert $). The sum in 
 Eq.(\ref{2.3b}) is over a complete set of reciprocal lattice vectors (RLV) $\vG$ with
 the property that $ e^{i\vG.\vR_{i}} = 1 $ for all $\vG$ and for all $\vR_{i}$. We refer 
 the first term of Eq.(\ref{2.3a}) as symmetry conserved and the second as symmetry broken 
 parts of single particle distribution \rhor.
 
 The two-particle density distribution $ \rho^{(2)}(\vr_{1},\vr_{2}) $ which gives probability of finding
 simultaneously a particle in volume element $ d\vr_{1} $ at $ \vr_{1} $ and a second particle in volume 
 element $ d\vr_{2} $ at $ \vr_{2} $, is defined as
  
\be
\rho^{(2)}(\vr_{1},\vr_{2}) = \left\langle \sum_{j}\sum_{k\neq j} \delta \left( \vr_{1} - \vr_{j} \right) 
\delta \left( \vr_{2} - \vr_{k} \right)\right\rangle . \label{2.4}
\ee

 The pair correlation function $ g(\vr_{1},\vr_{2}) $ is related to $\rho^{(2)}(\vr_{1},\vr_{2})$ by the
 relation,

\be
g(\vr_{1},\vr_{2}) = \frac{\rho^{(2)}(\vr_{1},\vr_{2})}{\rho(\vr_{1})\rho(\vr_{2})}. \label{2.5}
\ee

 The DPCF $ c(\vr_{1},\vr_{2}) $, which appears in the expression of free-energy 
 functional $ A[\rho] $ is related to the total pair correlation function $ h(\vr_{1},\vr_{2}) = g(\vr_{1},
 \vr_{2}) - 1 $  through the Ornstien - Zernike (OZ) equation \cite{2}
 
\be
c(\vr_{1},\vr_{2}) = h(\vr_{1},\vr_{2}) - \int d\vr_{3} c(\vr_{1},\vr_{3}) \rho(\vr_{3})h(\vr_{2},\vr_{3}).
\label{2.6}
\ee

 The second functional derivative of $ A[\rho] $ is expressed in terms of $c(\vr_{1},\vr_{2})$ as \cite{2}
 
\be
 \frac{\delta^{2}A[\rho]}{\delta\rho(\vr_{1})\ \delta\rho(\vr_{2})} =
  \frac{\delta(\vr_{1} - \vr_{2})}{\rho(\vr_{1})} - c(\vr_{1},\vr_{2}) \ , \label{2.7}
\ee
where $ \delta $ is Dirac function. The first term on the right hand side of this
 equation corresponds to ideal part $ A_{id}[\rho] $ of the 
 free energy whereas the second term corresponds to excess part $ A_{ex}[\rho] $ arising due to
  interparticle interactions.
   
 In a normal liquid all pair correlation functions defined above are simple function of number density $ \rho $ 
 and depend only on magnitude of interparticle separation $ \vert \vr_{2} - \vr_{1}\vert = r $. 
 This simplification is due to homogeneity which implies continuous translational symmetry and
  isotropy which implies continuous rotational symmetry. In a crystal which is both inhomogeneous 
  and anisotropic, pair correlation functions can be written as a sum of two terms; one that 
  preserves the continuous symmetry of the liquid and the other that breaks it \cite{13,16}. Thus
 
\begin{align}
h(\vr_{1},\vr_{2}) = h^{(0)}(\vert\vr_{2}-\vr_{1}\vert,\rho_{0}) + h^{(b)}(\vr_{1},\vr_{2};[\rho]) \label{2.8}
\\
c(\vr_{1},\vr_{2}) = c^{(0)}(\vert\vr_{2}-\vr_{1}\vert,\rho_{0}) + c^{(b)}(\vr_{1},\vr_{2};[\rho]). \label{2.9}
\end{align}

 While the symmetry conserving part ($ h^{(0)} $ and $ c^{(0)} $) depends on the magnitude of interparticle 
 separation $ r $  and is a function of average density $ \rho_{0} $, the symmetry broken parts $ h^{(b)} $ and 
 $ c^{(b)} $ are functional of \rhor (indicated by square bracket) and are invariant only under a discrete set 
 of translations corresponding to lattice vectors $ \vR_{i} $, 

\begin{align}
h^{(b)}(\vr_{1},\vr_{2}) = h^{b}(\vr_{1} + \vR_{i},\vr_{2} + \vR_{i}) \label{2.10} \\
c^{(b)}(\vr_{1},\vr_{2}) = c^{b}(\vr_{1} + \vR_{i},\vr_{2} + \vR_{i}) \label{2.11}
\end{align}

 If one chooses a centre of mass variable $ \vr_{c} = \left( \vr_{1} + \vr_{2}\right) \slash 2 $ and a 
 difference variable $ \vr = \vr_{2} - \vr_{1} $, then  one can see from Eqs. (\ref{2.10}) and (\ref{2.11}) that 
 $ h^{(b)} $ and $ c^{(b)} $ are periodic functions of the centre of mass variable and a continuous function of 
 the  difference variable \cite{18}. Thus

\begin{align}
h^{(b)}(\vr_{1},\vr_{2}) = \sum_{G} e^{i\vG . \vr_{c}} h^{(G)}(\vr) \label{2.12} \\
c^{(b)}(\vr_{1},\vr_{2}) = \sum_{G} e^{i\vG . \vr_{c}} c^{(G)}(\vr) \label{2.13}
\end{align}

 Since  $ h^{(G)} $ and  $ c^{(G)} $ are real and symmetric with respect to interchange of $ \vr_{1} $ and 
 $ \vr_{2} $;   $ h^{(-G)}(\vr) = h^{(G)} (\vr) $ and  $ h^{(G)}(-\vr) = h^{(G)}(\vr) $ and similar relations   
 holds for  $ c^{(G)}(\vr) $.

 Substitution of values of $ h(\vr_{1},\vr_{2}) $ and $ c(\vr_{1},\vr_{2}) $ given by Eqs. (\ref{2.8}) and 
 (\ref{2.9}) in Eq. (\ref{2.6}) allows us to split the OZ equation into two equations; one that contains
 $ h^{(0)} $,  $ c^{(0)} $ and \rhos while the other contains  $ h^{(b)} $,  $ c^{(b)} $ and 
 $ \rho(\vr_{3}) $ along with  $ h^{(0)} $,  $ c^{(0)} $ and \rhos :
 
 \be 
 h^{(0)}(\vert\vr_{2}-\vr_{1}\vert) = c^{(0)}(\vert\vr_{2}-\vr_{1}\vert) + \rho_{0} \int d\vr_{3} 
 c^{(0)}(\vert\vr_{3}-\vr_{1}\vert)  h^{(0)}(\vert\vr_{3}-\vr_{2}\vert) \label{2.14}
 \ee
 and

\begin{align}
 & h^{(b)}(\vr_{1},\vr_{2}) =  c^{(b)}(\vr_{1},\vr_{2}) +  \int d\vr_{3} c^{(0)}(\vert\vr_{3}-\vr_{1}\vert)
 (\rho(\vr_{3})-\rho_{0})  h^{(0)}(\vert\vr_{3}-\vr_{2}\vert) \nonumber \\
  &\qquad + \int d\vr_{3} \rho(\vr_{3})\left[ c^{(b)}(\vr_{1},\vr_{3}) h^{(0)}(\vert\vr_{3}-\vr_{2}\vert) +  
  c^{(0)}(\vert \vr_{3}-\vr_{2}\vert) h^{(b)}(\vr_{1},\vr_{3}) \right.\nonumber \\
  &\qquad \left. + c^{(b)}(\vr_{1},\vr_{3}) h^{(b)}(\vr_{1},\vr_{3}) \right]. \label{2.15} 
\end{align}  

 Eq. (\ref{2.14}) is the well known OZ equation of normal liquids. We use it along with a closure relation to 
 calculate the values of these correlation functions and their derivatives with respect to density \rhos. The 
 derivatives of $ c^{(0)}(r) $ are used to find values of three- and four- bodies direct correlation functions 
 of the isotropic phase.

 Eq. (\ref{2.15}) is the OZ equation for symmetry broken part of correlation functions. In order to make use of 
 it to find values of $ h^{(b)} $ and $ c^{(b)} $ for a given \rhor we need one more relation (closure relation) 
 that connects $ h^{(b)} $ with $ c^{(b)} $. Alternatively, if we know values of one of these functions then Eq. 
 (\ref{2.15}) can be used to find values of the other function \cite{19}. Here we calculate $c^{(b)}(\vr_{1},\vr_{2})$ 
 using a series in ascending powers of $(\rho(\vr)-\rho_{0})$.
   
\subsection{Calculation of $ h^{(0)} $, $ c^{(0)} $ and their derivatives with respect to $\rho$ }
 
 We use the OZ equation (\ref{2.14}) and a closure relation proposed by Roger and Young \cite{20} which mixes the 
 Percus-Yevick (PY) relation and the hypernetted chain (HNC) relation in such a way that at $ r = 0 $ 
 it reduces to the PY relation and for $ r\rightarrow \infty $ it reduces to the HNC relation and 
 is written as 
 
\begin{align} 
 h^{(0)}(r) = exp[-\beta u(r)]\left[ 1+\frac{exp[\chi(r) f(r)]}{f(r)}\right] - 1 \label{2.16}
\end{align}
 
  where $ \chi (r) = h^{(0)}(r) - c^{(0)}(r) $ and $ f(r) = 1 - exp(-\psi r) $ is a mixing function with 
  adjustable parameter $ 0\leq\psi\leq \infty $, to calculate pair correlation functions and their derivatives 
  with respect to density $ \rho $. The value of $ \psi $ is chosen to guarantee thermodynamic consistency 
  between the virial and compressibility routes to the equation of state \cite{20}.

  The differentiation of Eqs. (\ref{2.14}) and (\ref{2.16}) with respect to $ \rho $ yields the following 
  relations, 

\begin{align}
\frac{\partial h^{(0)}(r)}{\partial \rho} &= \frac{\partial c^{(0)}(r)}{\partial \rho} + \int{d\vr' c^{(0)}(r') 
h^{(0)}(\vert\vr'-\vr\vert)} \nonumber\\
 & \qquad + \rho \int{d\vr' \frac{\partial c^{(0)}(r')}{\partial \rho} h^{(0)}(\vert\vr'-\vr\vert)} \nonumber \\
 & \qquad + \rho \int{d\vr' c^{(0)}(r') \frac{\partial h^{(0)}(\vert\vr'-\vr\vert)}{\partial \rho}} \label{2.17}
\end{align} 
 and
\begin{align}
 \frac{\partial h^{(0)}(r)}{\partial \rho}  = exp[-\beta u(r)] {exp[\chi(r) f(r)]}
 \frac{\partial {\chi(r)}}{\partial \rho}, \label{2.18}
\end{align}
  
\begin{align}
\frac{\partial^{2} h^{(0)}(r)}{{\partial \rho}^{2}} &= \frac{\partial^{2} c^{(0)}(r)}{{\partial \rho}^{2}} + 2
 \int{d\vr'\left[ \frac {\partial c^{(0)}(r')} {\partial \rho} h^{(0)}(\vert\vr'-\vr\vert)
 + c^{(0)}(r') \frac{\partial h^{(0)}(\vert\vr'-\vr\vert)}{\partial \rho} \right]}  \nonumber \\
 & \qquad + \rho \int d\vr' \left[ 2 \frac{\partial c^{(0)}(r')}{\partial \rho} 
 \frac {\partial h^{(0)}(\vert\vr'-\vr\vert)}{\partial \rho}
  + c^{(0)}(r') \frac{\partial^{2} h^{(0)}(\vert\vr'-\vr\vert)}{{\partial \rho}^{2}} \right.  \nonumber \\
 & \qquad \qquad \qquad \qquad \left. + \frac{\partial^{2} c^{(0)}(r')}{{\partial \rho}^{2}} 
 h^{(0)}(\vert\vr'-\vr\vert) \right]   \label{2.19}
\end{align}

and

\begin{align}
\frac{\partial^{2} h^{(0)}(r)}{{\partial \rho}^{2}} = exp[-\beta u(r)] {exp[\chi(r) f(r)]}
\left[ \frac{\partial^{2} {\chi(r)}}{{\partial \rho}^{2}} + \left( \frac{\partial {\chi(r)}}
{\partial \rho} \right)^{2} f(r) \right] . \label{2.20}
\end{align}

 The solution of the closed set of coupled equations (\ref{2.14}) and (\ref{2.17})-(\ref{2.20}) gives values of
  $ h^{(0)}(r) $, $ c^{(0)}(r) $, $ \frac{\partial h^{(0)}(r)}{\partial \rho} $, $ \frac{\partial c^{(0)}(r)}
  {\partial \rho} $, $ \frac{\partial^{2} h^{(0)}(r)}{{\partial \rho}^{2}} $ and $ \frac{\partial^{2} c^{(0)}
  (r)}{{\partial \rho}^{2}} $ as a function of $r$ for a given potential $ u(r) $.

 The pair potential taken here are the inverse power potentials, $ u(r) = \epsilon \left(\sigma/r\right)^{n} $
 where $ \epsilon $, $ \sigma $ and n are potential parameters and $ r $ is the molecular separation.
 The parameter $ n $ measures softness of the potential; $ n = \infty $ corresponds to hard-sphere and
 $ n = 1 $ to the one component plasma. The reason for our choosing these potentials is that the
  range of potential can be varied by changing the value of $ n $ and the fact that the equation
  of state and melting curves of these potentials have been extensively investigated by computer
  simulations \cite{21,22,23,24,25,26,27,28} for several values of $ n $ so that "exact''
  results are available for comparison.
  The more repulsive $ \left( n \geq 7 \right) $ systems have been found to freeze into a 
  face-centred cubic (fcc) structure while the soft repulsions $ n < 7 $ freeze into a body-centred
  cubic crystal (bcc) structure. The fluid-bcc-fcc triple point is found to occur at
  $ \frac{1}{n} \simeq 0.15 $ \cite{25,26,28}. The atomic arrangements in the two cubic structures are very
  different; the fcc is close-packed in real space and the density inhomogeneity is much sharper
  than for the bcc which is open structure in real space but close-packed in Fourier space. However,
   in spite of this difference in the atomic arrangements, the two structures have small difference 
   in free energy (or chemical potential) at the fluid - solid transition \cite{25,26,27,28}
   and therefore a correct
    description of the relative stability of the two cubic structures is a stringent test for any theory.

 The inverse power potentials are known to have a simple scaling  property according to which the
  reduced thermodynamic properties depend on a single variable which is defined as
  
  \be 
  \gamma = \rho \sigma^{3} \left( \beta \epsilon \right)^{3/n} = \rho^{*} {T^{*}}^{(-3/n)}
  \nonumber
  \ee
  
  where $ \beta = \left. 1/ k_{B}T \right. $; $ k_{B} $ is the Boltzmann constant and T temperature.
  Using the scaling relation the potential is written as 
  
  \be 
  \beta u(r) = \left(\frac{4\pi}{3} \gamma \right)^{n/3} \frac{1}{r^{n}} \nonumber
  \ee 

  where $ r $ is measured in unit of $ a_{0}=\left(\frac{3}{4 \pi \rho}\right)^{1/3} $.

 In Fig. 1 we plot values of $ c^{(0)}(r) $, $ \frac{\partial c^{(0)}(r)}{\partial \rho} $ and 
 $ \frac{\partial^{2} c^{(0)}(r)}{{\partial \rho}^{2}} $ for $ n = 6 $ and $ \gamma = 2.30 $ which is close to 
 the freezing point.

\subsection{Calculation of three- and four-body direct correlation functions}

 The higher-body direct correlation function is related to the derivatives $ {\partial^{m} c^{(0)}(r)}/
 {{\partial \rho}^{m}} $ as follows \cite{2}:
 
\begin{align}
 \frac{\partial c^{(0)}(r,\rho_{0})}{\partial \rho} = \int {d\vr_{3} } \qquad 
 {c_3^{(0)}(\vr_{1},\vr_{2},\vr_{3};\rho_{0})}, \label{2.21}
\end{align}

\begin{align}
\frac{\partial^{2} c^{(0)}(r,\rho_{0})}{{\partial \rho}^{2}} &= \int {d\vr_{3} }\qquad 
\frac {c_3^{(0)}(\vr_{1},\vr_{2},\vr_{3};\rho_{0})} {\partial \rho} \nonumber \\
&= \int {d\vr_{3} \int {d\vr_{4} \qquad {c_4^{(0)}(\vr_{1},\vr_{2},\vr_{3},\vr_{4};\rho_{0})}}}, \label{2.22}
\end{align}

 etc., where $ c_m^{(0)} $ are m-body direct correlation function of the isotropic phase of density \rhos.
  These equations can be solved to find values of $ c_m^{(0)} $ by writing
   them as a product of pair functions. For $ c_3^{(0)}(\vr_{1},\vr_{2},\vr_{3}) $ 
   one can write as \cite{6},

 \begin{align}
  c_3^{(0)}(\vr_{1},\vr_{2},\vr_{3})&= t(r_{12})t(r_{13})t(r_{23}), \nonumber
 \end{align}
 \begin{align}
  & \equiv \nonumber  
 \end{align}
\vspace{-2.5cm}
 \begin{align}
 {\hspace{2.1cm}\includegraphics[width=1.5cm,height=1.5cm]{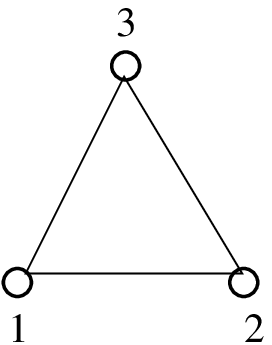}} \label{2.23}
  \end{align} 

 where a line linking particles $i$ and $j$ denotes a $ t(r) $ function and each circle ( representing a 
 particle) carry weight unity. The value of $ t(r) $ is found from the relation (\ref{2.21}),

 \begin{align}
 \frac{\partial c^{(0)}(r,\rho_{0})}{\partial \rho}=& \qquad \qquad \qquad , \nonumber
 \end{align}
 \vspace{-2.5cm}
 \begin{align}
 \hspace{1.5in}{\includegraphics[width=1.5cm,height=1.5cm]{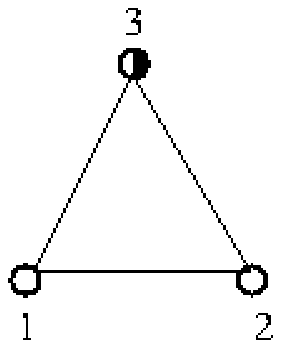}}
 \label{2.24} 
 \end{align}

 where the half -black circle represents the particle over which integration is performed
  over its all configurations and all circles carry 
 weight unity. Using known values of ${\partial c^{(0)}(r,\rho_{0})}/{\partial \rho_{0}} $ we solve this 
 equation to find values of $ t(r) $ for different density \rhos (or $ \gamma $)
 following a method outlined in ref \cite{6}. The values of
  $ t(r) $ as a function of $ r $ are shown in Fig. {\ref{Fig-tr}} for $ n=6, 4 $ and
  $ \gamma = 2.30, 5.60, $ respectively .
 
 Taking derivative of both sides of Eq.(\ref{2.23}) with respect to \rhos \  one gets,
  
\begin{align}
 \frac{\partial c_3^{(0)}(\vr_{1},\vr_{2},\vr_{3})}{\partial \rho_{0}} =  \frac{\partial t(r_{12})}
 {\partial \rho_{0}} t(r_{13})t(r_{23}) + t(r_{12})\frac{\partial t(r_{13})}{\partial \rho_{0}} t(r_{23}) +  
 t(r_{12})t(r_{13})\frac{\partial t(r_{23})}{\partial \rho_{0}}. \label{2.25}
\end{align}

 Substitution of this in Eq.(\ref{2.22}) leads to
 
\begin{align}
 \frac{\partial^{2} c^{(0)}(r)}{{\partial \rho_{0}}^{2}} = \int d\vr' \left[\frac{\partial t(r)}
 {\partial \rho_{0}} t(r')t(\vert \vr' - \vr \vert) + t(r)\frac{\partial t(r')}{\partial \rho_{0}} 
 t(\vert \vr' - \vr \vert) + t(r)t(r')\frac{\partial t(\vert \vr' - \vr \vert)}{\partial \rho_{0}}\right], 
  \label{2.26}
\end{align}

 where $ r_{12} = r $, $ r_{13} = r' $ and $ r_{23} = \vert \vr' - \vr \vert $. As values of $ t(r) $ are 
 known, Eq.(\ref{2.26}) is used to find values of $ {\partial t(r)}/{\partial \rho_{0}} $ in same way as 
 Eq.(\ref{2.24}) was used to find values of $ t(r) $. In Fig. {\ref{Fig-dtr}} we plot $ {\partial t(r)}/
 {\partial \rho_{0}} $ for $ n=4,6 $ and $ \gamma = 5.60, 2.30 $.
 
 Guided by the relation of Eq.(\ref{2.24}) we write $ {\partial t(r)}/{\partial \rho_{0}} $ as

\begin{align}
 \frac{\partial t(r)}{\partial \rho} &= s(r) \int d\vr'' s(r'')s(\vert \vr'' - \vr \vert), \nonumber \\
 \nonumber \\
  &\equiv \nonumber
\end{align}
 \vspace{-2.5cm}
 \begin{align}
 & {\hspace{1.0 cm}\includegraphics[width=1.5cm,height=1.5cm]{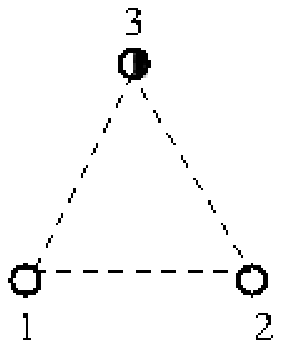}} \label{2.27} 
 \end{align}

 where a dashed line connecting particles \textit{i} and  \textit{j} is $ s(r) $ 
 function. Using the already determined values of 
 $ {\partial t(r)}/{\partial \rho_{0}} $ at a given value of $ \rho_{0} $ (or $ \gamma $)
 we determine values of $ s(r) $ in same way as values of $ t(r) $
 were determined from known values of 
 $ {\partial c^{(0)}(r)}/{\partial \rho_{0}} $. In Fig. {\ref{Fig-sr}} we plot values of 
 $ s(r) $ for $ n=6, 4 $ and $ \gamma = 2.30, 5.60 $ as a function of $ r $.
 
 From Eqs.(\ref{2.22}), (\ref{2.25}) and (\ref{2.27}) we get
 
 \begin{align}
 \nonumber \\
 c_4^{(0)}(\vr_{1},\vr_{2},\vr_{3},\vr_{4}) &= \qquad \qquad +\qquad \qquad +\qquad \qquad ,\nonumber 
 \end{align}
 \vspace{-2.5cm}
 \begin{align}
 {\hspace{0.1cm}\includegraphics[width=1.5cm,height=1.5cm]{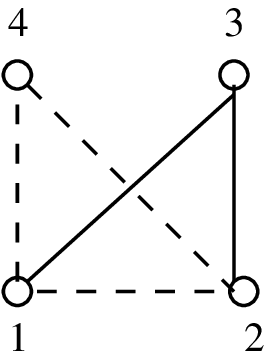}}   \nonumber
 \end{align}
 \vspace{-3.1cm}
 \begin{align}
 {\hspace{1.5 in}\includegraphics[width=1.5cm,height=1.5cm]{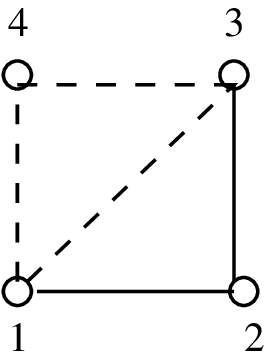}}   \nonumber
 \end{align}
  \vspace{-3.1cm}
 \begin{align}
 {\hspace{3.0 in}\includegraphics[width=1.5cm,height=1.5cm]{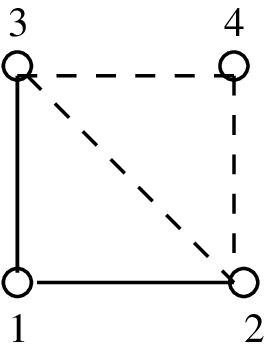}}  \label{2.28} 
 \end{align}

 where a dashed line represents $s(r)$-bond and a full line $t(r)$-bond. We calculate
  values of $ c_3^{(0)} $ and $ c_4^{(0)} $ and plot them in Appendix.
 
\subsection{Evaluation of $ c^{(b)}(\vr_{1},\vr_{2} $) }

 The function $ c^{(b)}(\vr_{1},\vr_{2}) $ can be expanded in ascending powers of $ (\rho(\vr)-\rho_{0}) $ 
 as \cite{2,13},
 
 \bse
 \begin{align}
 c^{(b)}(\vr_{1},\vr_{2};[\rho]) &= \int d\vr_{3} c_3^{(0)}(\vr_{1},\vr_{2},\vr_{3};\rho_{0})
  (\rho(\vr_{3})-\rho_{0}) \nonumber \\
  & \qquad + \frac{1}{2} \int d\vr_{3} \int d\vr_{3} c_4^{(0)}(\vr_{1},\vr_{2},\vr_{3},\vr_{4};\rho_{0})
  (\rho(\vr_{3})-\rho_{0})(\rho(\vr_{4})-\rho_{0}) \nonumber \\
  & \qquad + ...\qquad , \label{2.29a}    
 \end{align}
  
 \begin{align}
   &\equiv \qquad \qquad + \frac{1}{2} \qquad \qquad +\frac{1}{2} \qquad \qquad +\frac{1}{2} \qquad \qquad 
   + ..
 \label{2.29b}
 \end{align}
 \vspace{-2.4cm}
 \begin{align}
 {\hspace{-1.7in}\includegraphics[width=1.5cm,height=1.5cm]{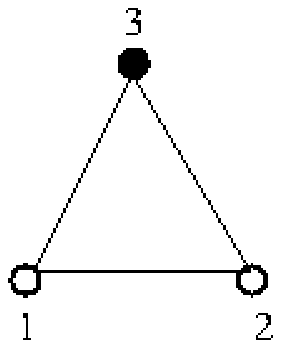}}   \nonumber
 \end{align}
 \vspace{-3.2cm}
 \begin{align}
 {\hspace{-0.8in}\includegraphics[width=1.5cm,height=1.5cm]{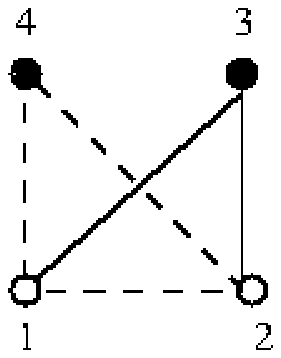}}   \nonumber
 \end{align}
\vspace{-1.2in}
 \begin{align}
 {\hspace{0.8 in}\includegraphics[width=1.5cm,height=1.5cm]{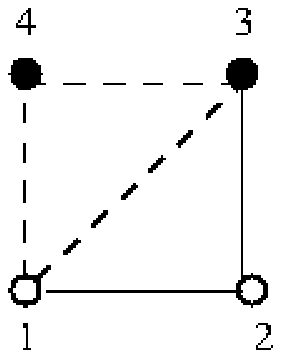}}   \nonumber
 \end{align}
  \vspace{-1.2in}
 \begin{align}
 {\hspace{2.8 in}\includegraphics[width=1.5cm,height=1.5cm]{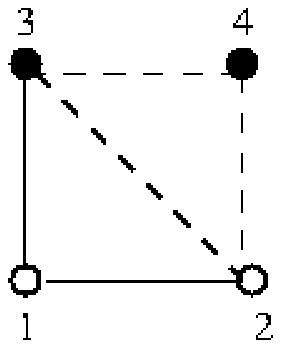}}  \nonumber  
  \end{align}    
 \ese

 where black circles represent integration over all configurations of these 
 particles and each carries 
 weight $ \rho(\vr_{i})-\rho_{0} = \sum_{G} \rho_{G} e^{i\vG.\vr_{i}} $ 
 whereas each white circle carries weight 
 unity. In writing Eq.(\ref{2.29b}) use has been made of Eqs.(\ref{2.23}) and (\ref{2.28}).
 
 Usefulness of series of Eq.(2.29) depends on how fast it converges and on our ability of 
 finding values of $ c_m^{(0)} $. We have already described the 
 calculation of $ c_3^{(0)} $ and $ c_4^{(0)} $.
 The same procedure can be used to find $ c_m^{(0)} $ for $ m > 4 $. We,
 however, find that for a wide range of 
 potentials it is enough to consider the first two terms of the series (2.29).
  In fact, for most potentials 
 representing the inter-particle interactions in real systems one may need to consider the first term only as 
 contribution made by the second term to the grand thermodynamic potential at the freezing point turns out 
 to be negligibly small unless the potential has a long range tail.

\subsubsection{Evaluation of first term of Eq.(2.29)}

 Substituting value of $(\rho(\vr_{3})-\rho_{0})$ from Eq.(2.3) and using notations, 
 $ \vr = \vr_{2} - \vr_{1} $, $ \vr' = \vr_{3} - \vr_{1} $, $ \vr_{c} = \frac{1}{2}(\vr_{1} + \vr_{2}) $
 we find

\begin{align}
 \qquad  \equiv c^{(b,1)}(\vr_{1},\vr_{2}) =  \sum_{G} \rho_{G} 
 e^{i \vG.\vr_{c}}t(r) e^{-\frac{1}{2}i \vG.\vr} \int d\vr' t(r') t(\vert\vr'-\vr\vert) e^{i \vG.\vr'}.
  \label{2.30}
\end{align}
\vspace{-2.5cm}
\begin{align}
{\hspace{-3.0in}\includegraphics[width=1.5cm,height=1.5cm]{3body11.eps}} \nonumber
\end{align}

This is solved to give \cite{13,14}

\begin{align}
 c^{(b,1)}(\vr_{1},\vr_{2}) = \sum_{G} e^{i \vG.\vr_{c}} \sum_{l m} c_{l}^{(G,1)}(r) Y_{lm}(\hat{r})
 Y_{lm}^{*}(\hat{G}), \label{2.31}
\end{align}

where

\begin{align}
c_{l}^{(G,1)}(r) = \rho_{G} \sum_{l_{1}}\sum_{l_{2}} \Lambda_{1}(l_{1},l_{2},l) j_{l_{2}}\left( \frac{1}
{2}Gr\right)B_{l_{1}}(r,G). \label{2.32}
\end{align}
 
 Here $ j_{l}(x) $ is the spherical Bessel function, $ Y_{lm}(\hat{x}) $ the spherical harmonics,
 
 \begin{align}
 \Lambda_{1}(l_{1},l_{2},l) = (i)^{l_{1}+l_{2}} (-1)^{l_{2}} \left[ \frac{(2l_{1}+1)(2l_{2}+1)}{(2l+1)}
 \right]^{\frac{1}{2}}\left[ C_{g}(l_{1},l_{2},l;0,0,0)\right]^{2}, \label{2.33}
 \end{align}
 
 and
 
 \begin{align}
 B_{l_{1}}(r,G) = 8 t(r) \int dk k^{2} t(k) j_{l_{1}}(kr) \int dr' r'^{2} t(r') j_{l_{1}}(kr') j_{l_{1}}(Gr'), 
 \label{2.34}
 \end{align}
 
 where $ C_{g} $ is the Clebsch-Gordan coefficient.
 The crystal symmetry dictates that $ l $ and $ l_{1}+l_{2} $ are even and for a cubic crystal, $m=0,\pm 4$.

 The values of $ c_{l}^{(G,1)}(r) $ depend on order parameters $ \rho_{G} = \rho_{0}\ \mu_{G} $, 
 where $ \mu_{G} = e^{-G^{2}/4\alpha} $ and on magnitude of $ \vG $. In Figs. \ref{Fig-cgr-3-1},
 \ref{Fig-cgr-3-2} we plot and compare values of
 $ c_{l}^{(G,1)}(r) $ for bcc and fcc crystals at the melting point for potential $ n = 6,\ \gamma_{s} = 2.32,
 \ \alpha_{bcc} = 18 $ and $ \alpha_{fcc} = 32 $ (see Table \ref{Tab1}).
 The values given in these figures are for the
 first and second sets of RLV's.  As expected, the values are far from negligible and differ considerably for
 the two structures. The value is found to decrease rapidly as the value of $ l $ is increased; the maximum
 contribution comes from $ l = 0 $. We also find, as shown in Fig. \ref{Fig-cgr-3-comp} ,
 the value of $ c_{l}^{(G,1)}(r) $ decreases rapidly
 as the magnitude of $ \vG $ vector increases; the maximum contribution comes from the first 
 two sets of RLV's. The other point to be noted is that at a given point $r$, values of $ c_{l}^{(G,1)}(r) $
 are positive for some $ \vG $ vectors while for others the values are negative leading to mutual cancellation
 in a quantity where summation over $ \vG $ is involved. 
 
\subsubsection{Evaluation of second term of Eq.(2.29)}

 The contribution arising from the second term of Eq.(2.29) is sum of three diagrams in which the last two
 contributions are equal. Thus,
 
 \begin{align}
 c^{(b,2)}(\vr_{1},\vr_{2}) = \frac{1}{2} \qquad \qquad \quad  + \qquad \qquad \qquad \qquad .
  \label{2.35}
 \end{align}
\vspace{-2.5cm}
\begin{align}
\hspace{-0.8cm} \includegraphics[width=1.5cm,height=1.5cm]{4body11.eps} \nonumber
\end{align}
\vspace{-3.1cm}
\begin{align}
\hspace{1.7in} \includegraphics[width=1.5cm,height=1.5cm]{4body21.eps}  \nonumber
\end{align}
 If we write $ \vr'' = \vr_{4}-\vr_{1} $ and $ \vr_{4} = \vr'' + \vr_{c} - \frac{1}{2} \vr $ and use other 
 notations defined above, the first diagram can be written as

 \begin{align}
  \frac{1}{2}\qquad \qquad \qquad & \equiv c^{(b,2,1)}(\vr_{1},\vr_{2})
 = \frac{1}{2} s(r) \sum_{G_{1}}\sum_{G_{2}} \rho_{G_{1}} \rho_{G_{2}} e^{i(\vG_{1}+\vG_{2}).(\vr_{c}-
 \frac{1}{2}\vr)} \nonumber \\
 & \int d\vr' t(r') t(\vert\vr'-\vr\vert) e^{i\vG_{1}.\vr'}
 \int d\vr'' s(r'') s(\vert\vr''-\vr\vert) e^{i\vG_{2}.\vr''}. \label{2.36}
 \end{align}  
   
\vspace{-3.7cm}
\begin{align}
\hspace{-2.5in} \includegraphics[width=1.5cm,height=1.5cm]{4body11.eps}  \nonumber
\end{align}
 
\vspace{2.0cm}
 This is solved to give
 
\begin{align}
 c^{(b,2,1)}(\vr_{1},\vr_{2}) = \sum_{G} e^{i\vG.\vr_{c}} \sum_{lm} \sum_{l'm'} c_{lm,l'm'}^{(G,2,1)}(r) 
 Y_{l'm'}^{*}(\hat{G}) Y_{lm}(\hat{r}), \label{2.37}
\end{align}

 where
 
\begin{align}
 c_{lm,l'm'}^{(G,2,1)}(r) = \sum_{G_{1}} \rho_{G_{1}} \rho_{K} \sum_{l_{1}m_{1}} \sum_{l_{2}m_{2}}
 \Lambda_{m m' m_{1} m_{2}}^{l l' l_{1} l_{2}} & M_{l_{1}}(r,G_{1}) M_{l_{2}}(r,K) \nonumber \\
 & j_{l'}\left(\frac{1}{2}Gr\right) Y_{l_{1}m_{1}}^{*}(\hat{G_{1}}) Y_{l_{2}m_{2}}^{*}(\hat{K}). \label{2.38}
\end{align}

 \qquad \qquad Here $ \roarrow K = \vG - \vG_{1} $,
 
\begin{align}
& \Lambda_{m m' m_{1} m_{2}}^{l l' l_{1} l_{2}} = 16 \sum_{l_{3}m_{3}}(i)^{l_{1}+l_{2}+l'} (-1)^{l'}
 \left[\frac{(2l_{1}+1)(2l_{2}+1)(2l'+1)}{(2l+1)} \right]^{1/2}  \nonumber \\
 &\qquad C_{g}(l_{1},l_{2},l_{3};0,0,0) C_{g}(l',l_{3},l;0,0,0) C_{g}(l_{1},l_{2},l_{3};m_{1},m_{2},m_{3})
  C_{g}(l',l_{3},l;m',m_{3},m); \label{2.39}  
\end{align} 
 
\begin{align}
 M_{l_{1}}(r,G_{1}) = \int dr' r'^{2} j_{l_{1}}(Gr') t(r') \int dk k^{2} t(k) j_{l_{1}}(kr) j_{l_{1}}(kr')
 \label{2.40} 
\end{align}
 and
\begin{align}
  M_{l_{2}}(r,K) = \int dr'' r''^{2} j_{l_{2}}(Kr'') s(r'') \int dk k^{2} s(k) j_{l_{2}}(kr) j_{l_{2}}(kr'')
  \qquad . \label{2.41} 
\end{align}

 The crystal symmetry dictates that all $ l_{i} $ are even and for a cubic crystal all $ m_{i} $ are $ 0 $ and
 $ \pm 4 $.
 
 From the second diagram of Eq.(2.35) we get

 \begin{align}
  \qquad \equiv & c^{(b,2,2)}(\vr_{1},\vr_{2}) =  t(r) \sum_{G_{1}}\sum_{G_{2}} 
 \rho_{G_{1}} \rho_{G_{2}} e^{i(\vG_{1}+\vG_{2}).(\vr_{c}-\frac{1}{2}\vr)} \nonumber \\
  & \qquad \int d\vr' s(r') t(\vert\vr'-\vr\vert) e^{i\vG_{1}.\vr'}
  \int d\vr'' s(r'') s(\vert\vr''-\vr'\vert) e^{i\vG_{1}.\vr''}. \label{2.42}
 \end{align} 
    
\vspace{-1.5in}
\begin{align}
\hspace{-3.0in} \includegraphics[width=1.5cm,height=1.5cm]{4body21.eps}  \nonumber
\end{align}
 
\vspace{2.0cm}

This is solved to give  
 
\begin{align}
 c^{(b,2,2)}(\vr_{1},\vr_{2}) = \sum_{G} e^{i\vG.\vr_{c}} \sum_{lm} \sum_{l'm'} c_{lm,l'm'}^{(G,2,2)}(r) 
 Y_{l'm'}^{*}(\hat{G}) Y_{lm}(\hat{r}), \label{2.43}
\end{align}

 where
 
\begin{align}
 c_{lm,l'm'}^{(G,2,2)}(r) = \sum_{G_{1}} \rho_{G_{1}} \rho_{K} \sum_{l_{1}m_{1}} \sum_{l_{2}m_{2}} 
 \sum_{l_{3}m_{3}} \Lambda_{mm'm_{1}m_{2}m_{3}}^{ll'l_{1}l_{2}l_{3}} 
 & N_{l_{1},l_{2},l_{3}}(r,G,G_{1}) \nonumber \\
 & j_{l'} \left(\frac{1}{2}Gr\right) Y_{l_{1}m_{1}}^{*}(\hat{G_{1}}) Y_{l_{2}m_{2}}^{*}(\hat{K}), \label{2.44}
\end{align}
 
\begin{align}
\Lambda_{mm'm_{1}m_{2}m_{3}}^{ll'l_{1}l_{2}l_{3}} &= 32 (i)^{l_{1}+l_{2}+l'} (-1)^{l'}
 \left[\frac{(2l_{1}+1)(2l_{2}+1)(2l'+1)}{(2l+1)} \right]^{1/2} \nonumber \\
& C_{g}(l_{1},l_{2},l_{3};0,0,0) C_{g}(l',l_{3},l;0,0,0) C_{g}(l_{1},l_{2},l_{3};m_{1},m_{2},m_{3}) 
 C_{g}(l',l_{3},l;m',m_{3},m); \label{2.45}  
\end{align} 
 
\begin{align}
 N_{l_{1},l_{2},l_{3}}(r,G,G_{1}) = t(r) \int dr' r'^{2} s(r') j_{l_{1}}(G_{1}r') B_{l_{2}}(r',K) 
 A_{l_{3}}(r,r'),
 \label{2.46} 
\end{align}

\begin{align}
 A_{l_{3}}(r,r') = \int dk k^{2} t(k) j_{l_{3}}(kr) j_{l_{3}}(kr')
 \label{2.47} 
\end{align} 
 and
\begin{align}
  B_{l_{2}}(r',K) = \int dr'' r''^{2} j_{l_{2}}(Kr'') s(r'') \int dk k^{2} s(k) j_{l_{2}}(kr') j_{l_{2}}(kr'').
 \label{2.48} 
\end{align}

 The total contribution arising from the second term of Eq.(2.29) is

\begin{align}
  c^{(b,2)}(\vr_{1}, \vr_{2}) = \sum_{G} e^{i\vG.\vr_{c}} \sum_{lm}\sum_{l'm'} \left[ c_{lml'm'}^{(G,2,1)}(r)
 +  c_{lml'm'}^{(G,2,2)}(r) \right] Y_{lm}(\hat r) Y_{l'm'}^{*}(\hat G) , \label{2.49}
\end{align}

 where $ l, l' $ are even and $ m = 0,\pm 4 $ for cubic lattices. In Figs. \ref{Fig-cgr-4-1} and \ref{Fig-cgr-4-2}
 we plot values of 

\begin{align}
  c_{lml'm'}^{(G,2)}(r) = c_{lml'm'}^{(G,2,1)}(r) + c_{lml'm'}^{(G,2,2)}(r) , \label{2.50}
\end{align}

 as a function of $r$ for bcc and fcc structures for $ n=6, \gamma_{s}=2.32, 
 \alpha_{bcc}=18 $ and $ \alpha_{fcc}=32 $.
 The values given in these figures are for the first two sets of RLV's for $ l=l'=0 $ and $ 2 $, and $ m=m'=0 $.
 These are the terms which mostly contribute to $ c_{lml'm'}^{(G,2)}(r) $; the contributions from terms
 $ l \neq l' $ and $ m \neq m' $ are approximately order of magnitude smaller. 
 For a bcc lattice we find two sets of values; one for $ \vG $
 vectors lying in the x-y plane and the other for the rest of the vectors. Since all vectors of the first set of 
 RLV's of a fcc lattice are out of x-y plane we get only one set of values. For the second set of RLV's of a
 fcc lattice, though two sets of values are found but they are close unlike the case of
  bcc lattice where the two
 sets of values differ not only in magnitude but also in sign. The values differ considerably for the two
 cubic structures. The value of $ c_{lml'm'}^{(G,2)}(r) $ decreases for both bcc and fcc structures 
 rapidly as the magnitude of $\vG$ vectors
 increases as was found in the case of $c_{l}^{(G,1)}(r)$. Furthermore, the values of $ c_{lml'm'}^{(G,2)}(r) $
 at a given value of $ r $ is positive for some $ \vG $ vector and negative for others.

 In order to compare magnitude of contributions made by the first and second terms of Eq.(2.29) we
 calculate ${\hat{c}}^{(G,1)}(k,\theta_{k},\phi_{k}) $ and $ {\hat{c}}^{(G,2)}(k,\theta_{k},\phi_{k}) $ 
 defined as 

 \begin{align}
 {\hat{c}}^{(G,1)}(k,\theta_{k},\phi_{k}) &= \rho \sum_{lm} \int d\vr c_{l}^{(G,1)}(r) e^{i \vec{k}.\vec{r}}
 Y_{lm}(\hat r) Y_{lm}^{*}(\hat G)   \nonumber \\
 &= 4\pi\ \rho \sum_{lm}\ (i)^{l}Y_{lm}(\hat k) Y_{lm}^{*}(\hat G) \int_{0}^{\infty} dr\ r^{2} 
 c_{l}^{(G,1)}(r) j_{l}(kr)
 \label{2.51}
\end{align}

 and
 
\begin{align}
 {\hat{c}}^{(G,2)}(k,\theta_{k},\phi_{k}) &= \rho \sum_{lm} \sum_{l'm'}\int d\vr c_{lml'm'}^{(G,2)}(r)
 e^{i \vec{k}.\vec{r}} Y_{lm}(\hat r) Y_{l'm'}^{*}(\hat G)   \nonumber \\
 &= 4\pi\ \rho \sum_{lm} \sum_{l'm'}\ (i)^{l}Y_{lm}(\hat k) Y_{l'm'}^{*}(\hat G) \int_{0}^{\infty} dr\ r^{2} 
 c_{lml'm'}^{(G,2)}(r) j_{l}(kr).
 \label{2.52}
\end{align}

 In Figs. \ref{Fig-fcgkang} and \ref{Fig-bcgkang} we compare using colour codes (shown on the
 right hand side of each figure) the values of these functions arising from the first and second terms
 of Eq.(2.29) for both fcc and bcc structures for $n=6$, \gams\ = 2.32, $\alpha_{fcc} = 32$
 and $\alpha_{bcc} = 18$. The values given in Fig. \ref{Fig-fcgkang} are for a fcc lattice 
 for a $\vG$ vector of first and a vector of second sets, i.e. $G_{1}a_{0} = 4.25$, 
 $\theta_{G_{1}} = {54.7\degree}$ and  $\phi_{G_{1}} = {45\degree}$ and $G_{2}a_{0} = 4.91$,
 $\theta_{G_{2}} = {0\degree}$,  $\phi_{G_{2}} = {90\degree}$. The values of $ka_{0}$
 are taken equal to 4.25, 4.91 which are magnitude of $G_{1}a_{0}$ and $G_{2}a_{0}$,
 respectively. In Fig. \ref{Fig-bcgkang} we compare the values of $ c^{(G,1)}(k,\theta_{k},\phi_{k}) $
 and $ c^{(G,2)}(k,\theta_{k},\phi_{k}) $ for a bcc lattice for $G_{1}a_{0} = 4.37$, 
 $\theta_{G_{1}} = {90\degree}$,  $\phi_{G_{1}} = {45\degree}$ and $G_{2}a_{0} = 6.19$,
 $\theta_{G_{2}} = {90\degree}$,  $\phi_{G_{2}} = {0\degree}$ and $ka_{0}$ = 4.37 and 6.19.
 From these figures it is clear that the contribution made by the second term to 
 $c^{(b)}(\vr_{1},\vr_{2})$ is small compared to the first term indicating fast convergence
 of the series. As is shown below, in the expression of free energy functional,
  $c^{(b)}(\vr_{1},\vr_{2})$ is averaged over density
 and order parameters and also there is  summation over $\vG$ vectors, As a consequence,
 the contribution of second term of Eq.(2.29) 
 is found to be order of magnitude smaller than
 the first term. We show that the consideration of the first two terms of Eq.(2.29)
 is enough to give accurate description of freezing transitions for a wide class
 of potentials.
 
\section{Free-energy functional and liquid-solid transition}

 The reduced free-energy functional $ A[\rho] $ of a symmetry broken phase can be written
 as \cite{13,14,17}

\begin{align}
 A[\rho] = A_{id}[\rho] + A_{ex}^{(0)}[\rho] + A_{ex}^{(b)}[\rho]  \label{3.1}
\end{align}

where
\begin{align}
 A_{id}[\rho] = \int d\vr \rho(\vr) \left[ ln(\rho(\vr)\Lambda)-1\right]   \label{3.2}
\end{align} 

\begin{align} 
 A_{ex}^{(0)}[\rho] &= A_{ex}(\rho_{l}) + \beta(\mu 
  - ln(\rho_{l}\Lambda)) \int d\vr (\rho(\vr)-\rho_{l}) \nonumber \\ 
 & - \frac{1}{2} \int d\vr_{1} \int d\vr_{2} (\rho(\vr_{1})-\rho_{l})(\rho(\vr_{2})-\rho_{l}) 
  {\overline{c}}^{(0)}(\vert\vr_{2}-\vr_{1}\vert)  \label{3.3}
\end{align}

and
\begin{align}
 A_{ex}^{(b)}[\rho] = - \frac{1}{2} \int d\vr_{1} \int d\vr_{2} (\rho(\vr_{1})-\rho_{0})
  (\rho(\vr_{2})-\rho_{0}){\overline{c}}^{(b)}(\vr_{1},\vr_{2}) \label{3.4}
\end{align}

 Here $ \Lambda $ is cube of the thermal wavelength associated with a molecule, $ \beta = 
 (k_{B}T)^{-1} $, $ k_{B} $ being the Boltzmann constant and $ T $ is the temperature,
 $ A_{ex}^{(0)}(\rho_{l}) $ is excess reduced free energy of the  coexisting isotropic liquid
 of density \rhol\ and chemical potential $ \mu $ and $ \rho_{0} = \rho_{l} \left( 1 + \Delta
  \rho^{*} \right) $ is the average density of the solid. 
 
\begin{align}
 {\overline{c}}^{(0)}(\vert\vr_{2}-\vr_{1}\vert) = 2 \int_{0}^{1} d\lambda \lambda \int_{0}^{1} d\lambda^{'}
 c^{(0)}\left( \vert\vr_{2}-\vr_{1}\vert;\rho_{l} + \lambda \lambda^{'}(\rho_{0}-\rho_{l})\right)  
 \label{3.5} 
\end{align}

 and
 
\begin{align}
 {\overline{c}}^{(b)}(\vr_{1},\vr_{2}) = 4 \int_{0}^{1} d\lambda \lambda \int_{0}^{1} d\lambda^{'}
 \int_{0}^{1} d\xi \xi \int_{0}^{1} d\xi^{'}
 c^{(b)}\left( \vr_{1},\vr_{2};\lambda \lambda^{'} \rho_{0},\xi \xi^{'} \rho_{G} \right)\quad .  
 \label{3.6} 
\end{align}

 The expression for the symmetry conserving part of reduced excess free energy $ A_{ex}^{(0)}[\rho] $ given by 
 Eq.(\ref{3.3}) is found by performing double functional integration of \cite{13,17}

\begin{align}
 \dfrac{\delta^{2}A_{ex}^{(0)}[\rho]}{\delta \rho(\vr_{1})\ \delta \rho(\vr_{2})} = 
 - {c}^{(0)}(\vert\vr_{2}-\vr_{1}\vert).  \label{3.7}
\end{align}
 
 This integration is carried out in the density space taking the coexisting uniform fluid of density \rhol\ 
 and chemical potential $ \mu $ as a reference. The expression for the symmetry broken part $ A_{ex}^{(b)}[\rho] $
 given by Eq.(\ref{3.4}) is found by performing double functional integration of \cite{13,17}

\begin{align}
 \dfrac{\delta^{2}A_{ex}^{(b)}[\rho]}{\delta \rho(\vr_{1})\ \delta \rho(\vr_{2})} = 
 - {c}^{(b)}(\vr_{1},\vr_{2}),  \label{3.8}
\end{align}

 in the density space corresponding to the symmetry broken phase. The path of integration in this space
 is characterized by two parameters $ \lambda $ and $ \xi $. These parameters vary from 0 to 1. The parameter
 $ \lambda $ raises the density from zero to the final value \rhos\ as it varies from 0 to 1, whereas parameter
 $ \xi $ raises the order parameter from 0 to its final value $ \rho_{G} $. The result is independent of 
 the order of integration. 
 
 In locating the transition the grand thermodynamic potential defined as
\begin{align}
 - W = A - \beta \mu \int d\vr \rho(\vr) \label{3.9} 
\end{align}
 is generally used as it ensures the pressure and chemical potential of both phases remain equal at the 
 transition. The transition point is determined by the condition $ \Delta W = W_{l} - W = 0 $, where $ W_{l} $
 is the grand thermodynamic potential of the co-existing liquid. The expression of $ \Delta W $ is found to be
 \cite{13,14}
 
\begin{align}
 \Delta W &= \int d\vr \left[ \rho(\vr) ln\left(\frac{\rho(\vr)}{\rho_{l}}\right)
  -\left(\rho(\vr)-\rho_{l}\right)\right] \nonumber \\
 & \qquad - \frac{1}{2} \int d\vr_{1} \int d\vr_{2} (\rho(\vr_{1})-\rho_{l})(\rho(\vr_{2})-\rho_{l}) 
  {\overline{c}}^{(0)}(\vert\vr_{2}-\vr_{1}\vert) \nonumber \\
 & \qquad - \frac{1}{2} \int d\vr_{1} \int d\vr_{2} (\rho(\vr_{1})-\rho_{0})(\rho(\vr_{2})-\rho_{0}) 
  {\overline{c}}^{(b)}(\vr_{1},\vr_{2}). \label{3.10}
\end{align}

 Minimization of $ \Delta W $ with respect to \rhor subject to the perfect crystal constraint leads to

\begin{align}
 ln\frac{\rho(\vr_{1})}{\rho_{l}} = \phi +  \int d\vr_{2} (\rho(\vr_{2})-\rho_{l}) 
  \tilde{c}^{(0)}(\vert\vr_{2}-\vr_{1}\vert) + \int d\vr_{2} (\rho(\vr_{2})-\rho_{0}) 
  \tilde{c}^{(b)}(\vr_{1},\vr_{2}), \label{3.11}
\end{align}

where
 
\begin{align}
 {\tilde{c}}^{(0)}(\vert\vr_{2}-\vr_{1}\vert) = \int_{0}^{1} d\lambda 
 c^{(0)}\left( \vert\vr_{2}-\vr_{1}\vert; \rho_{l} + \lambda (\rho_{0}-\rho_{l})\right)   \nonumber
\end{align}

 and
 
\begin{align}
 {\tilde{c}}^{(b)}(\vr_{1},\vr_{2}) =  \int_{0}^{1} d\lambda  \int_{0}^{1} d\xi c^{(b)}\left( \vr_{1},\vr_{2}; 
 \lambda \rho_{0}, \xi \rho_{G})\right)   \nonumber
\end{align}

 The value of the Lagrange multiplier $ \phi $ in Eq.(\ref{3.11}) is found from the condition
 
\begin{align}
\frac{1}{V} \int d\vr \frac{\rho(\vr)}{\rho_{0}} = 1  \label{3.12}
\end{align}
 where V is volume of the system.

 It may be noted that, in principle, one needs only values of symmetry conserved and symmetry broken parts
 of the DPCF to determine \rhor that minimizes the grand potential $ W $. In practice, however, it is 
 found convenient to do minimization with respect to an assumed form of \rhor. The ideal part is calculated using 
 a form of \rhor which is a superposition of normalized Gaussians centred around the lattice sites,
 
\begin{align}
\rho(\vr) = \left(\frac{\alpha}{\pi}\right)^{3/2} \sum_{n} exp\left[-\alpha\left(\vr-\vR_{i}\right)^{2}\right] 
\label{3.13}
\end{align}

 where $ \alpha $ is the variational parameter that characterizes the width of the Gaussian; the square root of
 $ \alpha $ is inversely proportional to the width of a peak. It thus measures the non-uniformity; 
 $ \alpha = 0 $ corresponds to the limit of a uniform liquid and an increasing
  value of $ \alpha $ corresponds to
 increasing localization of particles on their respective lattice sites defined by vectors $ \vR_{i} $. For the 
 interaction part it is convenient to use the expression of \rhor given by Eq.(2.3). The Fourier transform
 of Eq.(\ref{3.13}) leads to $ \rho_{G} = \rho_{0} \mu_{G} $, where $ \mu_{G} = e^{-G^{2}/4\alpha} $.
 
\subsection{Evaluation of $ {\overline{c}}^{(0)}(r) $ and $ {\overline{c}}^{(b)}(\vr_{1},\vr_{2}) $}

 The values of $ {\overline{c}}^{(0)}(r) $ for a given liquid density \rhol\ and the average crystal density \rhos\
 are found from the known values of $ c^{(0)}(r,\rho) $ where $ \rho $ varies from \rhol\ to \rhos\ by performing 
 integrations in Eq.(\ref{3.5}) which can be rewritten as

\begin{align}
 {\overline{c}}^{(0)}(r,\rho_{0}) = 2 \int_{0}^{1} d\lambda \lambda \int_{0}^{1} d\lambda^{'}
 c^{(0)}\left( r;\rho_{l}(1 + \lambda \lambda^{'}\Delta\rho^{*})\right)  
 \label{3.14} 
\end{align}

 where $ \Delta \rho^{*} = (\rho_{0} - \rho_{l})/\rho_{l} $. The integrations have been done numerically using
 a very fine grid for variables $ \lambda $ and $ \lambda^{'} $ . Since at the freezing point $ \rho_{l}
 \Delta\rho^{*} << 1$ one can use Taylor expansion to solve Eq.(\ref{3.14}) leading to 
 
\begin{align}
{\overline{c}}^{(0)}(r,\rho_{0}) = c^{(0)}(r,\rho_{l}) + \frac{1}{3}\rho_{l}\Delta\rho^{*}
\frac{\partial c^{(0)}(r,\rho_{l})}{\partial \rho_{l}} + O\left(\rho_{l}^{2}{\Delta\rho^{*}}^{2}\right) 
\label{3.15}
\end{align}

 Since the order parameters that appear in $ {\overline{c}}^{(b)}(\vr_{1},\vr_{2}) $ are linear in
 $ c^{(b,1)}(\vr_{1},\vr_{2}) $ and quadratic in $ c^{(b,2)}(\vr_{1},\vr_{2}) $, the integration over $ \xi $
 variables in Eq.(\ref{3.4}) can be performed analytically leading to

\begin{align}
 {\overline{c}}^{(b)}(\vr_{1},\vr_{2}) &= \sum_{G} e^{i\vG.\vr_{c}} \left[
 \sum_{lm} {\overline{c}}_{l}^{(G,1)}(r) {Y}_{lm}^{*}(\hat{G})
  Y_{lm}(\hat{r}) \right. \nonumber \\
 & \qquad \left. + \sum_{lm}\sum_{l'm'} \left( {\overline{c}}_{lm,l'm'}^{(G,2,1)}(r)
 + {\overline{c}}_{lm,l'm'}^{(G,2,2)}(r) \right) {Y}_{l'm'}^{*}(\hat{G}) 
 Y_{lm}(\hat{r}) \right]   \label{3.16}
\end{align}

where
\begin{align}
{\overline{c}}_{l}^{(G,1)}(r) = \frac{1}{3} \rho_{G} \sum_{l_{1}} \sum_{l_{2}} \Lambda_{1}(l_{1},l_{2},l)
j_{l_{2}}\left(\frac{1}{2}Gr\right) \overline{B}_{l_{1}}(r,G) ,  \label{3.17}
\end{align}
 
\begin{align}
{\overline{c}}_{lm,l'm'}^{(G,2,1)}(r) = \frac{1}{6} \sum_{G_{1}} \rho_{G_{1}} \rho_{K} \sum_{l_{1}m_{1}}
 \sum_{l_{2}m_{2}} &\Lambda_{mm'm_{1}m_{2}}^{ll'l_{1}l_{2}} j_{l}\left(\frac{1}{2}Gr\right) \nonumber \\
& \qquad \overline{Q}_{l_{1}l_{2}}(r,G,G_{1}) {Y}_{l_{1}m_{1}}^{*}(\hat{G_{1}}){Y}_{l_{2}m_{2}}^{*}(\hat{K}),
\label{3.18}
\end{align}

\begin{align}
{\overline{c}}_{lm,l'm'}^{(G,2,2)}(r) = \frac{1}{6} \sum_{G_{1}} \rho_{G_{1}} \rho_{K} \sum_{l_{1}m_{1}}
 \sum_{l_{2}m_{2}} \sum_{l_{3}m_{3}} &\Lambda_{mm'm_{1}m_{2}m_{3}}^{ll'l_{1}l_{2}l_{3}}
j_{l'}\left(\frac{1}{2}Gr\right) \nonumber \\
& \qquad \overline{N}_{l_{1}l_{2}l_{3}}(r,G,G_{1}) {Y}_{l_{2}m_{2}}^{*}(\hat{G_{1}}){Y}_{l_{3}m_{3}}^{*}(\hat{K}),
\label{3.19}
\end{align}

with

\begin{align}
 {\overline{B}}_{l_{1}}(r,G) = 2 \int_{0}^{1} d\lambda \lambda \int_{0}^{1} d\lambda^{'}
 B_{l_{1}}\left( r,G; \lambda \lambda^{'}\rho\right), \label{3.20} 
\end{align}

\begin{align}
 {\overline{Q}}_{l_{1}l_{2}}(r,G,G_{1}) = 2 \int_{0}^{1} d\lambda \lambda \int_{0}^{1} d\lambda^{'}
 Q_{l_{1}l_{2}}\left( r,G,G_{1}; \lambda \lambda^{'}\rho\right), \label{3.21} 
\end{align}

\begin{align}
 Q_{l_{1}l_{2}}\left( r,G,G_{1};\rho\right) = M_{l_{1}}\left(r,G_{1};\rho\right) M_{l_{2}}\left(r,K;\rho\right),
  \nonumber
\end{align}

and

\begin{align}
 {\overline{N}}_{l_{1}l_{2}l_{3}}(r,G,G_{1}) = 2 \int_{0}^{1} d\lambda \lambda \int_{0}^{1} d\lambda^{'}
 N_{l_{1}l_{2}l_{3}}\left( r,G,G_{1}; \lambda \lambda^{'}\rho\right). \label{3.22} 
\end{align}

 The quantities $ B_{l_{1}}( r,G) $, $ M_{l_{1}}(r,G_{1}) $, $ M_{l_{2}}(r,K) $ and 
 $ N_{l_{1}l_{2}l_{3}}( r,G,G_{1}) $ are defined by Eqs. (\ref{2.34}), (\ref{2.40}), (\ref{2.41})
 and (\ref{2.46}) respectively. The integrations over $ \lambda $ and 
 $ \lambda^{'} $ have been performed numerically by varying them from 0 
 to 1 on a fine grid and evaluating the functions $ B_{l_{1}} $, $ Q_{l_{1}l_{2}} $ and $ N_{l_{1}l_{2}l_{3}} $
 on these densities. Since these functions vary smoothly with density and their values have been evaluated 
 at closely spaced values of density the result found for $ {\overline{c}}^{(b)}(\vr_{1},\vr_{2}) $ is expected
 to be accurate.
 
 \subsection{Evaluation of $ \Delta W $}
 
 Substituting expression of \rhor given by Eqs.(2.3) and (\ref{3.14}) and of $ {\overline{c}}^{(0)}(r) $
 and $ {\overline{c}}^{(b)}(\vr_{1},\vr_{2}) $ given above in Eq.(\ref{3.10}) we find
 
\begin{align}
\frac{\Delta W}{N} = \frac{\Delta W_{id}}{N} + \frac{\Delta W_{0}}{N} +\frac{\Delta W_{b}^{(1)}}{N} +
 \frac{{\Delta W}_{b}^{(2)}}{N}  \label{3.23}
\end{align}

 where
 
\begin{align}
\frac{\Delta W_{id}}{N} = 1 - (1 + \Delta \gamma) \left[\frac{5}{2} + ln \rho_{l} - 
\frac{3}{2} ln\left(\frac{\alpha}{\pi}\right)\right] \label{3.24}
\end{align}

\begin{align}
\frac{\Delta W_{0}}{N} = - \frac{1}{2} \Delta \gamma \widehat{\overline{c}}^{(0)}(0) -
\frac{1}{2} (1 + \Delta \gamma)^{2} \sum_{G \neq 0} {\vert\mu_{G}\vert}^{2} {\widehat{\overline{c}}}^{(0)}(G)
\label{3.25}
\end{align}

\begin{align}
\frac{\Delta W_{b}^{(1)}}{N} = -\frac{1}{2} \rho_{l}(1 + \Delta \gamma)^{2} {\sum_{G}}^{'} {\sum_{G_{2}}}^{'}
\mu_{G_{2}} \mu_{-G-G_{2}} \widehat{\overline{c}}^{(G,1)}\left(\vG_{2} + \frac{1}{2} \vG\right) \label{3.26}
\end{align}

\begin{align}
\frac{\Delta W_{b}^{(2)}}{N} = -\frac{1}{2} \rho_{l}(1 + \Delta \gamma)^{2} {\sum_{G}}^{'} {\sum_{G_{2}}}^{'}
\mu_{G_{2}} \mu_{-G-G_{2}} \widehat{\overline{c}}^{(G,2)}\left(\vG_{2} + \frac{1}{2} \vG\right) \label{3.27}
\end{align}

where $ \Delta \gamma = \left(\gamma_{s} - \gamma_{l}\right)/{\gamma_{l}} $; 
the subscripts $s$ and $l$ stand for solid 
and liquid, respectively. Here $ \Delta W_{id} $, $ \Delta W_{0} $, 
$ \Delta W_{b}^{(1)} $ and $ \Delta W_{b}^{(2)} $ are respectively,
 the ideal, symmetry conserving and symmetry broken contributions
from first and second terms of series (2.29) to $ \Delta W $.
 The prime on summation in Eqs.(\ref{3.26}),
(\ref{3.27}) indicates the condition $ \vG \neq 0 $, $ \vG_{1} \neq 0 $, $ \vG_{2} \neq 0 $, 
$ \vG+\vG_{1} \neq 0 $ and $ \vG+\vG_{2} \neq 0 $ and

\begin{align}
\widehat{\overline{c}}^{(0)}(G) = \int d\vr {\overline{c}}^{(0)}(r,\gamma_{l}) e^{i\vG.\vr} ,  \label{3.28}
\end{align}

\begin{align}
\widehat{\overline{c}}^{(G,1)}\left(\vG_{2} + \frac{1}{2}\vG\right) &= \frac{1}{3} \mu_{G} \sum_{l_{1}}
 \sum_{l_{2}} \Lambda_{1}(l_{1},l_{2},l) {Y}_{lm}^{*}(\hat{G}) \int d\vr\ j_{l_{2}}
 \left(\frac{1}{2}Gr\right) \nonumber \\
  &\qquad \overline{B}_{l_{1}}(r,G) e^{i\left(\vG_{2} + \frac{1}{2}\vG\right).\vr} 
   {Y}_{lm}(\hat{r}) , \label{3.29}
\end{align}

\begin{align}
\widehat{\overline{c}}^{(G,2)}\left(\vG_{2} + \frac{1}{2}\vG\right) =
 \widehat{\overline{c}}^{(G,2,1)}\left(\vG_{2} + \frac{1}{2}\vG\right) + 2 \   
 \widehat{\overline{c}}^{(G,2,2)}\left(\vG_{2} + \frac{1}{2}\vG\right), \nonumber
\end{align}

\begin{align}
\widehat{\overline{c}}^{(G,2,1)}\left(\vG_{2} + \frac{1}{2}\vG\right) &= \frac{1}{6} \sum_{G_{1}}
 \mu_{G_{1}} \mu_{K} \sum_{lm}\sum_{l'm'}\sum_{l_{1}m_{1}} \sum_{l_{2}m_{2}}
 \Lambda_{mm'm_{1}m_{2}}^{ll'l_{1}l_{2}} {Y}_{l'm'}^{*}(\hat{G})
 {Y}_{l_{1}m_{1}}^{*}(\hat{G_{1}}){Y}_{l_{2}m_{2}}^{*}(\hat{K})  \nonumber \\
 &\qquad \int d\vr\ j_{l_{2}}\left(\frac{1}{2}Gr\right)
  \overline{Q}_{l_{1}l_{2}}(r,G,G_{1}) e^{i\left(\vG_{2} + \frac{1}{2}\vG\right).\vr} 
 {Y}_{lm}(\hat{r}) ,
 \label{3.30}
\end{align}

\begin{align}
\widehat{\overline{c}}^{(G,2,2)}\left(\vG_{2} + \frac{1}{2}\vG\right) &= \frac{1}{6} \sum_{G_{1}} \mu_{G_{1}} 
\mu_{K} \sum_{lm}\sum_{l'm'}\sum_{l_{1}m_{1}} \sum_{l_{2}m_{2}} \sum_{l_{3}m_{3}} 
  \Lambda_{mm'm_{1}m_{2}m_{3}}^{ll'l_{1}l_{2}l_{3}} {Y}_{l'm'}^{*}(\hat{G})
 {Y}_{l_{2}m_{2}}^{*}(\hat{G_{1}}){Y}_{l_{3}m_{3}}^{*}(\hat{K}) \nonumber \\
  &\qquad \int d\vr\ j_{l_{2}}\left(\frac{1}{2}Gr\right) 
 \overline{N}_{l_{1}l_{2}l_{3}}(r,G,G_{1}) e^{i\left(\vG_{2} + \frac{1}{2}\vG\right).\vr} 
 {Y}_{lm}(\hat{r}) .
 \label{3.31}
\end{align}

 The terms $\frac{\Delta W_{0}}{N}, \frac{\Delta W_{b}^{(1)}}{N}$ and $ \frac{\Delta W_{b}^{(2)}}{N}$
 are respectively second, third and fourth orders in order parameters.

\section{Results for liquid-crystal transition}

We use above expression of $ {\Delta W}/N $ to locate the
 liquid - fcc crystal and the liquid - bcc 
crystal transitions by varying \gaml\ , \dga\ and $ \alpha $. For a given 
\gaml\  and \dga, $ {\Delta W}/N $
is minimised with respect to $ \alpha $; next \dga\  is varied untill the lowest value
 of $ {\Delta W}/N $
at its minimum is found. If this lowest value of $ {\Delta W}/N $ at its minimum 
is not zero, then \gaml\
is varied until $ {\Delta W}/N = 0 $. The values of transition parameters,
 \gaml, \dga\  and $ \alpha $ for
a given lattice structure can also be found from simultaneous solution of equations 
 $ \frac{\partial}{\partial(\Delta\gamma)} \left(\frac{\Delta W}{N}\right) = 0 $, 
 $ \frac{\partial}{\partial\alpha} \left(\frac{\Delta W}{N}\right) = 0 $
 and $ {\Delta W}/N = 0 $.

In Table \ref{Tab1} we compare values of different terms of
 $ {\Delta W}/N  $ (see Eq.(\ref{3.23})) at the freezing 
point for potentials with $ n = 4,6,6.5,7,12 $ and $ \infty $. 
The values corresponding to hard spheres are taken
 from ref.\cite{14}. The contribution made by the 
symmetry broken part to the grand thermodynamic potential
at the freezing point is substantial and its importance 
increases with the softness of the potential. For example,
while for $ n = \infty $ the contribution of the symmetry 
broken part is about $ 8\% $ of the contribution made by the symmetry conserved
part, it increases to $ 45\% $ for $ n = 4 $. As this
contribution is negative, it stabilizes the solid phase.
Without it the theory strongly overestimates the stability
of the fluid phase especially for softer potentials.
This explains why the Ramakrishanan - Yussouff theory 
gives good results for hard core potentials but fails
for potentials that have soft core and/or attractive tail.

The other point to be noted from these results is about the convergence of the series (2.29) which
has been used to calculate $ c^{(b)}(\vr_{1}, \vr_{2}) $. The contribution made by the second term
of the series to the grand thermodynamic potential at the freezing point
is found to be negligible compared to that of the first term for $ n \geq 6 $ and for $ n < 6 $,
though the contribution is small but not negligible. For example, while for $ n = 6 $ this contribution
is about $ 2\% $ of the first term, for $ n = 4 $ this increases to $ 18\%  $. From these results one can 
conclude that the first two terms of the series of Eq.(2.29) are enough to describe the freezing
transition for a wide class of potentials.

In Table \ref{Tab2}, we compare results of freezing parameters \gaml , \gams, \dga , 
 Lindemann parameter $ L_{n} $ and $ \dfrac{P\sigma^{3}}{\epsilon}$ 
 where P is the pressure at the transition point,
of the present calculation with those found from computer simulations \cite{21,22,23,24,25,26,27,28}
 and with the results found by others \cite{12,30,31,32} using
approximate free energy functionals. The Lindemann parameter is defined as the ratio of the mean field
displacement of a particle to the nearest neighbour distance
 in the crystal. For the fcc crystal with the Gaussian
density profile of Eq.(\ref{3.13}) it is given as

\begin{align}
L_{n} = \left(\frac{3}{a_{fcc}^{2}\alpha}\right)^{1/2} ,  \label{4.1}
\end{align}

where $ a_{fcc} = \left(4/\rho_{0}\right)^{1/3} $ is the fcc lattice constant. For the bcc crystal,

\begin{align}
L_{n} = \left(\frac{2}{a_{bcc}^{2}\alpha}\right)^{1/2} ,  \label{4.2}
\end{align}

where $ a_{bcc} = \left(2/\rho_{0}\right)^{1/3} $ is the bcc lattice constant.

 In Fig. \ref{Fig-TP} we plot \gaml\ vs  $ 1/n $ at the transition found from simulations and from
 the present calculations.

One may note that simulation results have spread (see Table \ref{Tab2})
 and do not agree within each others uncertainties. This 
may be due to application of different theoretical methods used in 
locating the transition and system sizes in 
the calculations. The other sources of errors include the existence of an interface, truncation of the
potential, free-energy bias, etc. Agrawal and Kofke \cite{25} who have reported
results for $ 0 \leq 1/n \leq 0.33 $ have
considered a system of 500 particles only. Since they have not 
used finite size corrections, their results
for softer potentials (say $ n\lesssim 6 $) may not be 
accurate. For example, they reported that for 
$ 1/n > 0.16 $ \ fluid freezes into a bcc structure 
but for $ 1/n = 0.25 $ they found that \gaml\ 
for the fluid - bcc transition is higher than that of 
the fluid-fcc transition. The recent calculations where large systems have been
considered \cite{26,27,28} results
are available for $ n \geq 5 $ (or $ 1/n < 0.2 $). From
 these results it is found that fluid freezes into 
fcc crystal for $ n \geq 7 $ and for $ n < 7 $ the bcc 
structure is preferred; the fluid-bcc-fcc triple
point is estimated to be close to $ 1/n \sim 0.15 $.

From Table \ref{Tab2} and Fig. \ref{Fig-TP} we find that our results are
in very good agreement with simulation results for all
cases. We find that for $ n > 6.5 $ the fluid freezes 
into fcc structure while for $ n \leq 6 $ it freezes into
bcc structure. The fluid-bcc-fcc triple point is found at $ \frac{1}{n} = 0.158 $
 (see the inset in Fig. \ref{Fig-TP}).
The value of Lindemann parameter found by us is, however, somewhat lower than those
found by Agrawal and Kofke \cite{25} and Saija et al \cite{29}.
 The energy difference between the two cubic structures at the transition is found to be
 small in agreement with the simulation results \cite{28}.
 
\section{Summary and Perspectives.} 

 We used a free energy functional for a crystal proposed by Singh and Singh \cite{13} to investigate the
 crystallization of fluids interacting via power law potentials. This free-energy functional was
 found by performing  double functional integration in the density space of a relation that relates
 the second functional derivative of $ A[\rho] $ with respect to \rhor\ to the DPCF of the crystal.
 The expression found for $ A[\rho] $ is exact and contains both the symmetry conserved part of
 the DPCF, $ c^{(0)}(r, \rho) $ and the symmetry broken part $ c^{(b)}(\vr_{1}, \vr_{2}) $. The 
 symmetry conserved part corresponds to the isotropy and homogeneity of the phase and passes smoothly
 to the frozen phase at the freezing point, whereas the symmetry broken part arises due to 
 heterogeneity which sets in at the freezing point and vanishes in the liquid phase. The values of
 $ c^{(0)}(r) $ and its derivatives with respect to density $ \rho $ as a function of interparticle
 separation $r$ have been determined using an integral equation theory comprising the OZ equation
 and the closer relation of Roger and Young \cite{20}. From the results of $ \frac{\partial c^{(0)}(r)}
 {\partial \rho} $ and $ \frac{\partial^{2} c^{(0)}(r)}{{\partial \rho}^{2}} $ we calculated 
 the three- and four-bodies direct correlation functions of the isotropic phase. These 
 results have been used in a series written in ascending powers of the order parameters to 
 calculate $ c^{(b)}(\vr_{1}, \vr_{2}) $. The contributions made by the first and second terms
 of the series have been calculated for bcc and fcc crystals. The contribution made by second term 
 is found to be considerably smaller than the first term indicating that the first two terms are
  enough to give accurate values for $ c^{(b)}(\vr_{1}, \vr_{2}) $. The values of 
  $ c^{(G)}(\vr) $ for bcc and fcc structures are found to differ considerably.
 
  The contribution of symmetry broken part of DPCF to the free energy is found to depend 
 on the nature of pair potentials; the contribution increases with softness of potentials.
 In case of power law potentials we found that the contribution to the grand thermodynamic
 potential at the freezing point arising from the second term of the series (2.29) which involves
 four-body direct correlation function is negligible for $n > 6$ and small but not negligible
 for $n < 6$. For $n = 4$ the contribution made by the second term is about $18\%$ of the first term.
 The contribution made by second term is positive whereas the contribution of the first term
 is negative. As the net contribution made by the symmetry broken term is negative, it
 stabilizes the solid phase. Without the inclusion of this term the theory strongly
 overestimates the stability of the fluid phase especially for softer potentials. Our results
 reported in this paper and elsewhere \cite{14,17} explain why the Ramakrishanan - Yussouff
 theory gives good results for hard core potentials but fails for potentials that have soft
 core /or attractive tail.
 
  The agreement between theory and simulation values of freezing parameters found
  for potentials with $n$ varying from $4$ to $\infty$ indicates
 that the free energy functional used here with values of $ c^{(b)}(\vr_{1}, \vr_{2}) $ calculated
 from the first two terms of the series (2.29) provides an accurate theory for freezing transitions for a 
 wide class of potentials. Since this free energy functional takes into account  the spontaneous
 symmetry breaking, it can be used to study various phenomena of ordered phases near their
 melting points.
 
 {\bf{\underline{Acknowledgements}}}: One of us (A.S.B.) thanks University Grants Commission 
 (New Delhi, India) for award of research fellowship.

\newpage
\appendix
\section{}

  In this appendix we calculate $ c_{3}^{(0)}(\vr_{1}, \vr_{2}, \vr_{3}) $ and 
 $ c_{4}^{(0)}(\vr_{1}, \vr_{2}, \vr_{3}, \vr_{4}) $.
     Using the notation $ r = |\vr_{2} - \vr_{1}| $, $ r' = |\vr_{3} - \vr_{1}| $
 and $|\vr' - \vr| = |\vr_{3} - \vr_{2}| $  we write $ c_{3}^{(0)}(\vr_{1}, \vr_{2}, \vr_{3}) $
 as (see Eq.(2.23))
 
 \begin{align}
  c_{3}^{(0)}(\vr, \vr') = t(r) t(r') t(|\vr'-\vr|).            \label{A1}
 \end{align}

 The function $t(|\vr'-\vr|)$ can be expanded in spherical harmonics,
 
 \begin{align}
  t(|\vr'-\vr|) = \dfrac{2}{\pi} \sum_{lm} A_{l}(r,r') {Y}_{lm}(\hat{r}){Y}_{lm}^{*}(\hat{r'}), \label{A2}
 \end{align}
 
 where,
 
 \begin{align}  
 A_{l}(r,r') = \int_{0}^{\infty}dq\ q^{2} t(q) j_{l}(qr) j_{l}(qr') .   \label{A3}
 \end{align}
 
 Here $j_{l}(x)$ is the spherical Bessel function and $Y_{lm}(\hat{r})$ the spherical harmonics.
 
 From Eqs.(\ref{A1}) and (\ref{A2}) we get
 
  \begin{align} 
 c_{3}^{(0)}(\vr, \vr') = \dfrac{2}{\pi} \sum_{lm} D_{l}(r,r') {Y}_{lm}(\hat{r})
 {Y}_{lm}^{*}(\hat{r'}),               \label{A4}
 \end{align}
 
 where
 
  \begin{align}
 D_{l}(r,r') = A_{l}(r,r') t(r) t(r') . \nonumber
 \end{align}
 
 The Fourier transform of Eq.(\ref{A4})defined as
  \begin{align}
  \hat{c}_{3}^{(0)}(\vq_{1},\vq_{2}) = \rho^{2} \int d\vr\ \int d\vr'\ 
  e^{-i\vq_{1}.\vr} e^{-i\vq_{2}.\vr'} c_{3}^{(0)}(\vr, \vr'), \nonumber
 \end{align}
 
 gives
 
  \begin{align}
 \hat{c}_{3}^{(0)}(\vq_{1},\vq_{2}) = 32\pi \sum_{lm} (-1)^{l} 
 D_{l}(q_{1},q_{2}) {Y}_{lm}(\hat{q_{1}}) {Y}_{lm}^{*}(\hat{q_{2}}),         \label{A5}
 \end{align}
 
 where,
 
  \begin{align}
  D_{l}(q_{1},q_{2}) = \rho^{2} \int dr\ r^{2}\int dr'\ r'^{2} j_{l}(q_{1}r)
  j_{l}(q_{2}r') D_{l}(r,r') .        \label{A6}
 \end{align}
 
 The value of $ \hat{c}_{3}^{(0)}(\vq_{1},\vq_{2}) $ is plotted in Fig. \ref{Fig-3bodycorr-1}
 for $q_{1} = q_{2} = q_{max}$ for various angle $\theta$ such that
 $0 < |\vq_{1}+\vq_{2}| < 2 q_{max}$, where $\theta$ is angle between $\vq_{1}$ and $\vq_{2}$
 as shown in the figure. The values plotted in this figure correspond to $ qa_{0}=4.3 $
 and for $ n=6,\ \gamma_{l}=2.30 $ (full line) and
  $ n=12,\ \gamma_{l}=1.17 $ (dashed line). In Fig. \ref{Fig-3bodycorr-2}
 we plot values of $ \hat{c}_{3}^{(0)}(\vq_{1},\vq_{2}) $ for equilateral triangle with various
 side lengths. The values for $ n=12, \gamma_{l}=1.17 $ are in good agreement with the values
 given in ref \cite{6} (see Figs. 3 and 4 of ref \cite{6}).
 
 For $ c_{4}^{(0)}(\vr_{1}, \vr_{2}, \vr_{3}, \vr_{4}) $ the contribution arises from
 three diagrams shown in Eq.(\ref{2.28}). Using the notation $ |\vr_{4}-\vr_{1}| = r'' $,
 $ |\vr_{4}-\vr_{2}| = |\vr''-\vr| $, $ |\vr_{4}-\vr_{3}| = |\vr''-\vr'| $ and other notations
  defined above we get
 
  \begin{align}
 c_4^{(0)}(\vr, \vr', \vr'') &= \qquad \qquad \qquad +\qquad \qquad +\qquad \qquad ,\nonumber 
 \end{align}
 \vspace{-2.5cm}
 \begin{align}
 {\hspace{0.1cm}\includegraphics[width=1.5cm,height=1.5cm]{4body1.eps}}   \nonumber
 \end{align}
 \vspace{-3.1cm}
 \begin{align}
 {\hspace{1.5 in}\includegraphics[width=1.5cm,height=1.5cm]{4body2.eps}}   \nonumber
 \end{align}
  \vspace{-3.1cm}
 \begin{align}
 {\hspace{3.0 in}\includegraphics[width=1.5cm,height=1.5cm]{4body3.eps}}  \label{A7}
 \end{align}
 
 Each diagram of Eq.(\ref{A7}) has two circles connected by three bonds- two $s$ - bonds (dashed line) and
 one $t$ - bond (full line), one of the remaining circles is connected by two $t$ - bonds and the other
 by two $s$ - bonds. By permuting circles one can convert one diagram into another. The values 
 of $ c_{4}^{(0)}(\vr, \vr', \vr'') $ depend on three vectors $ \vr, \vr' $ and $ \vr'' $.
 
 We calculate $ \hat{c}_{4}^{(0)}(\vq_{1}, \vq_{2}, \vq_{3}) $ defined as
 
  \begin{align}
   \hat{c}_{4}^{(0)}(\vq_{1}, \vq_{2}, \vq_{3}) = \rho^{3}\ \int d\vr\ \int d\vr'\ 
  \int d\vr''\ e^{-i\vq_{1}.\vr} e^{-i\vq_{2}.\vr'} e^{-i\vq_{2}.\vr'} c_{4}^{(0)}
  (\vr, \vr', \vr'') , \label{A8}
 \end{align}
 
 Using Eq.(\ref{A7}) and writing each diagram in terms of $t$ and $s$ bonds we get
  
 \begin{align}
 \hat{c}_{4}^{(0)}(\vq_{1}, \vq_{2}, \vq_{3}) &= \dfrac{108}{\pi^{2}} \sum_{l_{1}m_{1}}
 \sum_{l_{2}m_{2}} \sum_{l_{3}m_{3}} (-i)^{(l_{1}+l_{2}+l_{3})} 
 \Lambda_{l_{1}l_{2}l_{3}}^{m_{1}m_{2}m_{3}} M_{l_{1}l_{2}l_{3}}(q_{1}, q_{2}, q_{3})
 \nonumber \\
 & \qquad \left[{Y}_{l_{3}m_{3}}^{*}(\hat{q_{1}}) {Y}_{l_{1}m_{1}}(\hat{q_{2}}) 
 {Y}_{l_{2}m_{2}}(\hat{q_{3}}) + {Y}_{l_{1}m_{1}}(\hat{q_{1}}) 
 {Y}_{l_{3}m_{3}}^{*}(\hat{q_{2}}) {Y}_{l_{2}m_{2}}(\hat{q_{3}}) \right. \nonumber \\
 & \qquad \left. + (-1)^{l_{1}}\ {Y}_{l_{1}m_{1}}(\hat{q_{1}}) 
 {Y}_{l_{3}m_{3}}^{*}(\hat{q_{2}}) {Y}_{l_{2}m_{2}}(\hat{q_{3}})\right]  ,  \label{A9} 
 \end{align}
 
 where
 \begin{align}
 \Lambda_{l_{1}l_{2}l_{3}}^{m_{1}m_{2}m_{3}} = \left[\dfrac{(2l_{1}+1)(2l_{2}+1)}
 {4\pi(2l_{3}+1)}\right]^{1/2} C_{g}(l_{1},l_{2},l_{3};0,0,0) 
 C_{g}(l_{1},l_{2},l_{3};m_{1},m_{2},m_{3}),  \label{A10}
 \end{align}
 
 and
 
 \begin{align}
 M_{l_{1}l_{2}l_{3}}(q_{1}, q_{2}, q_{3}) &= \rho^{3}\ \int_{0}^{\infty} dr\ r^{2} 
  s(r) \int_{0}^{\infty} d\vr'\ r'^{2} t(r') \int_{0}^{\infty} d\vr''\ r''^{2}
 s(r'') \nonumber \\
 & j_{l_{3}}(q_{1}r) j_{l_{1}}(q_{2}r') j_{l_{2}}(q_{3}r'') 
 A_{l_{1}}(r,r') E_{l_{2}}(r,r'') .  \label{A11}
 \end{align}
 
$ A_{l_{1}}(r,r') $ is defined by Eq.(\ref{A3}). $ E_{l_{2}}(r,r'') $ is given as
 
 \begin{align}  
 E_{l}(r,r'') = \int_{0}^{\infty}dq\ q^{2} s(q) j_{l}(qr) j_{l}(qr'') .  \label{A12}
 \end{align}
 
 The values of $ \hat{c}_{4}^{(0)}(\vq_{1}, \vq_{2}, \vq_{3}) $ depend on magnitudes
 and directions of vectors $ \vq_{1} $, $ \vq_{2} $ and $ \vq_{3} $. In Figs. 
 \ref{Fig-4bodycorr-1} and \ref{Fig-4bodycorr-2} we use color codes(shown at
 the right hand side of each figure) to plot values of 
 $ \hat{c}_{4}^{(0)}(\vq_{1}, \vq_{2}, \vq_{3}) $
 for $ q_{1} = q_{2} = q_{3} = q_{max} $ as a function of $ \phi_{q_{2}} $ and
  $ \phi_{q_{3}} $ for different choices of $ \theta_{q_{2}} $ and $ \theta_{q_{3}} $.
  The values of $ {q_{max}a_{0}} $ is taken equal to 4.3 as in Fig. \ref{Fig-3bodycorr-1}.
   While the values plotted in Fig. \ref{Fig-4bodycorr-1} correspond to
    $ \theta_{q_{1}} = 0^{\circ} $, the values plotted in Fig. \ref{Fig-4bodycorr-2} correspond
     to $ \theta_{q_{1}} = 90^{\circ} $ and $ \phi_{q_{1}} = 0^{\circ} $.
  These figures show how the values of $\hat{c}_{4}^{(0)}(\vq_{1}, \vq_{2}, \vq_{3}) $
  depend on orientations of vectors $ \vq_{1} $, $ \vq_{2} $ and $ \vq_{3} $. Emergence of
  ordering in maxima and minima depending on orientations of these vectors is evident.
  


\newpage
\begin{center}
\begin{scriptsize}
\begin{table*}
\caption{Freezing parameters \gaml\ , \dga\ and
the contributions of ideal symmetry conserving and symmetry broken parts
arising from first and second terms of Eq.(2.29) to $ {\Delta W}/N $
at the transition point.}
\vspace{0.5cm}
\label{Tab1}
\begin{ruledtabular}
\begin{tabular}{|>{\scriptsize}c|>{\scriptsize}c|>{\scriptsize}c|>{\scriptsize}c|>{\scriptsize}c|>{\scriptsize}c|>{\scriptsize}c|>{\scriptsize}c|}
\small
$ n $&$ Lattice $&$\gamma_{l}$&$\Delta \gamma $&$ \Delta W_{id}/N $&$ \Delta W_{o}/N $&$ \Delta W_{b}^{(1)}/N $&$ \Delta W_{b}^{(2)}/N $ \\ \hline
$ 4 $& bcc &$ 5.57 $ &$ 0.007 $&$ 2.86 $&$ -2.09 $&$ -0.94 $&$ 0.17 $ \\
$ { } $& fcc &$ 5.60 $ &$ 0.008 $&$ 3.52 $&$ -2.64 $&$ -1.03 $&$ 0.15 $ \\ \hline
$ 6 $& bcc &$ 2.30 $ &$ 0.011 $&$ 2.56 $&$ -1.99 $&$ -0.58 $&$ 0.01 $ \\
$ { } $& fcc &$ 2.32 $ &$ 0.012 $&$ 3.48 $&$ -2.75 $&$ -0.72 $&$ 0.002 $ \\ \hline
$ 6.5 $& bcc &$ 2.04 $ &$ 0.014 $&$ 2.38 $&$ -1.89 $&$ -0.50 $&$ 0.001 $ \\
$ { } $& fcc &$ 2.03 $ &$ 0.013 $&$ 3.34 $&$ -2.69 $&$ -0.66 $&$ 0.001 $ \\ \hline
$ 7 $& bcc &$ 1.86 $ &$ 0.015 $&$ 2.29 $&$ -1.85 $&$ -0.44 $&$ 0.000 $ \\
$ { } $& fcc &$ 1.84 $ &$ 0.014 $&$ 3.39 $&$ -2.76 $&$ -0.63 $&$ 0.000 $ \\ \hline
$ 12 $& fcc &$ 1.17 $ &$ 0.034 $&$ 3.71 $&$ -3.14 $&$ -0.57 $&$ 0.000 $ \\ \hline
$ \infty $& fcc &$ 0.937 $ &$ 0.106 $&$ 4.44 $&$ -4.10 $&$ -0.34 $&$ 0.000 $ \\
\end{tabular}
\end{ruledtabular}
\end{table*}
\end{scriptsize}
\end{center}

\begin{center}
\LTcapwidth=\textwidth
\renewcommand{\arraystretch}{0.75}
\begin{longtable}{|>{\scriptsize}c|>{\scriptsize}c|>{\scriptsize}c|>{\scriptsize}c|>{\scriptsize}c|>{\scriptsize}c|>{\scriptsize}c|>{\scriptsize}c|}
\caption{{Comparison of parameters \gaml\ , \gams\ , \dga\ , the Lindemann parameter L and
 the pressure P at the coexistence found from
 different free-energy functional and computer simulations. MWDA stands for modified
 weighted density approximation, RY DFT stands for Ramakrishnan - Yussouff Density
  functional theory. MHNC stands for modified hypernetted-chain closure relation
  and MSMC for Mayer sampling Monte Carlo.}}\\[1mm]
\hline \hline
$ n $ & $ Lattice $ &  Theory/ Simulation  & $ \gamma_{l} $ & $ \gamma_{s} $ &
$ \Delta\gamma $ & $ L $ & $ \dfrac{P\sigma^{3}}{\epsilon} $\\[1mm]
\hline
\endfirsthead
\multicolumn{8}{c}%
{\tablename\ \thetable\ -- \scriptsize{\textit{Continued from previous page}}} \\[1mm]
\hline \hline
$ n $ & $ Lattice $ &  Theory/ Simulation  & $ \gamma_{l} $ & $ \gamma_{s} $ & 
$ \Delta\gamma $ & $ L $ & $ \dfrac{P\sigma^{3}}{\epsilon}$\\[1mm]
\hline
\endhead
\hline \hline \multicolumn{8}{c}{\scriptsize {\textit{Continued on next page}}} \\
\endfoot
\hline \hline \multicolumn{8}{c}{\scriptsize {\textit{\textbf{NOTE-}{* indicates values obtained from
interpolation of the tabulated values.}}}} \\
\endlastfoot 
\label{Tab2}
$ \infty $ & { fcc } & {Present result} & 0.937 & 1.036 & 0.106 & 0.09 & 11.46  \\
{ } & { } & {MWDA-static reference \cite{12}} & 0.863 & 0.964 & 0.115 & 0.13 & { } \\
{ } & { } & {MWDA \cite{32}} & 0.906 & 1.044 & 0.116 & 0.10 & { } \\
{ } & { } & {RY DFT \cite{30,31}} & 0.980 & 1.146 & 0.174 & 0.06 & { } \\
{ } & { } & {Simulation \cite{22}} & 0.939 & 1.037 & 0.104 & $ \sim 0.13 $ & { } \\
{ } & { } & {Simulation \cite{23}} & 0.942 & 1.041 & 0.105 & { } & { } \\
{ } & { } & {MC Simulation \cite{25}} & 0.94 & 1.041 & 0.107 & 0.12 & 11.70  \\
{ } & { } & {MC Simulation \cite{26}} & 0.939 & 1.037 & 0.104 & { } & 11.57  \\ \hline
 12 & { fcc } & {Present result} & 1.17 & 1.21 & 0.034 & 0.11 & 23.67  \\
{ } & { } & {MWDA-static reference \cite{12}} & 1.12 & 1.16 & 0.037 & 0.14 & { } \\
{ } & { } & {MWDA/MHNC \cite{32}} & 1.19 & 1.26 & 0.059 & 0.10 & { } \\
{ } & { } & {MC Simulation* \cite{25}} & 1.17 & 1.22 & 0.042 & 0.14 & 23.64  \\
{ } & { } & {MSMC technique \cite{27}} & 1.16 & 1.20 & 0.037 & { } & 23.24  \\
{ } & { } & {MC Simulation \cite{26}} & 1.16 & 1.21 & 0.037 & { } & 23.41  \\ \hline
 7  & { fcc } & {Present result} & 1.84 & 1.87 & 0.014 & 0.12 & 64.97  \\
{ } & { } & {MC Simulation* \cite{25}} & 1.85 & 1.88 & 0.017 & 0.15 & 64.98 \\
{ } & { } & {MC Simulation \cite{26}} & 1.84 & 1.87 & 0.016 & { } & 64.22  \\ 
{ } & { bcc } & {Present result} & 1.86 & 1.89 & 0.015 & 0.18 & 67.12  \\
{ } & { } & {MC Simulation \cite{26}} & 1.83 & 1.86 & 0.015 & { } & 63.88  \\ \hline
 6.5 & { fcc } & {Present result} & 2.03 & 2.06 & 0.013 & 0.12 & 80.11  \\
{ } & { } & {MC Simulation* \cite{25}} & 2.04 & 2.07 & 0.014 & 0.15 & 80.40  \\
{ } & { bcc } & {Present result} & 2.04 & 2.07 & 0.014 & 0.17 & 78.98  \\
{ } & { } & {MC Simulation* \cite{28,29}} & 2.03 & 2.05 & 0.010 & 0.18 & 78.40  \\ \hline
 6  & { fcc } & {Present result} & 2.32 & 2.35 & 0.012 & 0.12 & 103.7  \\
{ } & { } & {MWDA-static reference \cite{12}} & 2.33 & 2.35 & 0.007 & 0.17 & { } \\
{ } & { } & {MC Simulation* \cite{25}} & 2.34 & 2.37 & 0.012 & 0.15 & 104.5  \\
{ } & { } & {MC Simulation \cite{26}} & 2.32 & 2.35 & 0.012 & { } & 103.0  \\
{ } & { bcc } & {Present result} & 2.30 & 2.33 & 0.011 & 0.16 & 101.22  \\
{ } & { } & {MC Simulation* \cite{25}} & 2.32 & 2.35 & 0.011 & 0.17 & 103.6  \\
{ } & { } & {MSMC technique \cite{27}} & 2.30 & 2.32 & 0.011 & { } & 100.1  \\
{ } & { } & {MC Simulation \cite{26}} & 2.30 & 2.33 & 0.012 & { } & 100.0  \\
{ } & { } & {MC Simulation* \cite{28,29}} & 2.29 & 2.31 & 0.009 & 0.18 & 99.34  \\
\hline
 4  & { fcc } & {Present result} & 5.60 & 5.63 & 0.008 & 0.12 & 565.6  \\
{ } & { } & {MWDA-static reference \cite{12}} & 5.22 & 5.26 & 0.008 & 0.13 & { } \\
{ } & { } & {MC Simulation \cite{25}} & 5.68 & 5.71 & 0.005 & 0.17 & 637.0  \\
{ } & { bcc } & {Present result} & 5.57 & 5.61 & 0.007 & 0.16 & 561.2  \\
{ } & { } & {MWDA-static reference \cite{12}} & 5.05 & 5.09 & 0.008 & 0.18 & { } \\
{ } & { } & {MC Simulation \cite{25}} & 5.73 & 5.75 & 0.004 & 0.18 & 648.0  \\
\end{longtable}
\end{center}
\begin{figure}[h]
\vspace{0.8cm}
\includegraphics[height=4.0in,width=4.0in]{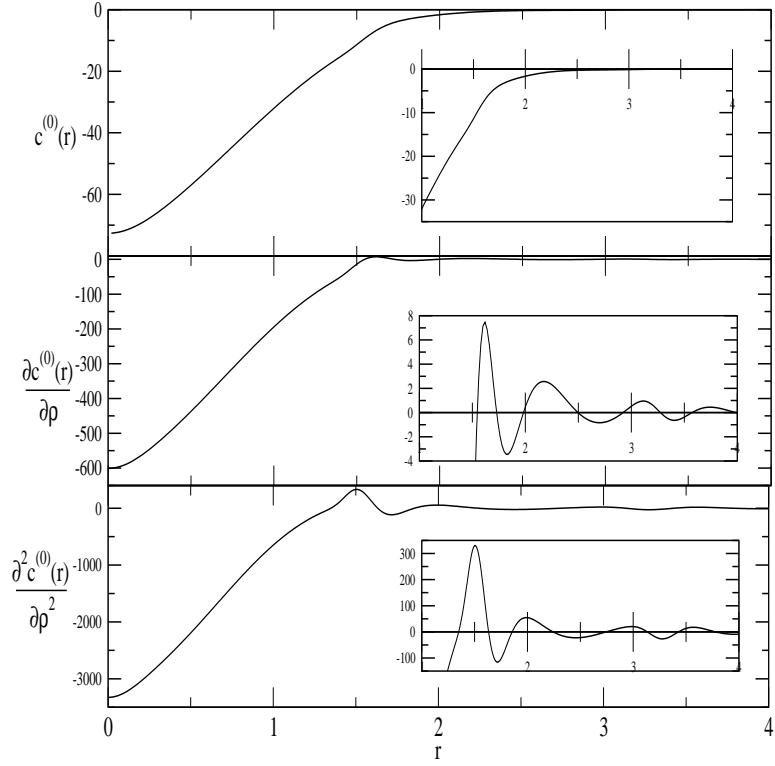}
\caption{Plots of $ c^{(0)}(r) $, $ \frac{\partial c^{(0)}(r)}{\partial \rho} $ and 
 $ \frac{\partial^{2} c^{(0)}(r)}{{\partial \rho}^{2}} $ vs $ r $ for $ n = 6 $ and 
 $ \gamma = 2.30 $ which is close to the freezing point. The distance $ r $ is in unit of 
$ a_{0}=\left(\frac{3}{4 \pi \rho}\right)^{1/3}$. Insets show magnified values of respective
quantities for $ r \geq 1 $.} \label{Fig-cr}
\end{figure}

\begin{figure}[h]
\vspace{0.8cm}
\includegraphics[height=3.0in,width=4.0in]{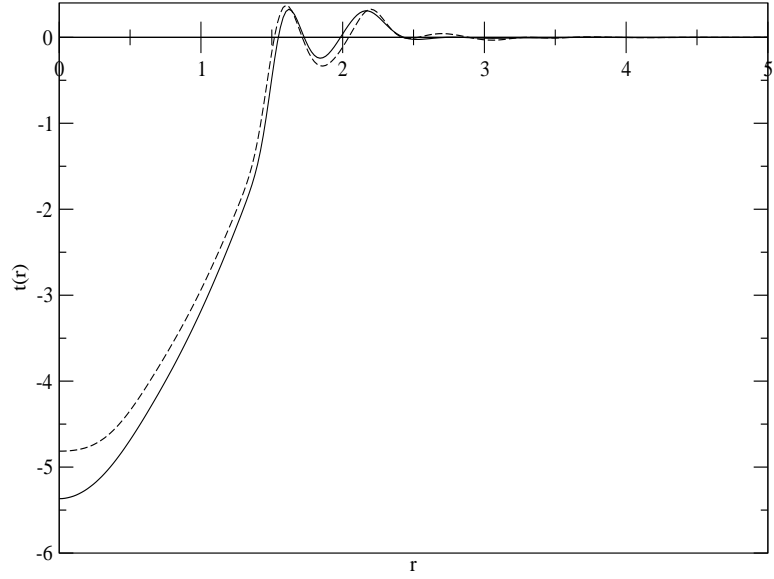}
\caption{Plot of $ t(r) $ vs $ r $ for $ n = 6 $ at $ \gamma = 2.32 $ 
and $ n = 4 $ at $ \gamma = 5.60 $. The distance $ r $ is in unit of 
$ a_{0}=\left(\frac{3}{4 \pi \rho}\right)^{1/3}$. The dashed curve represents 
values for $ n = 4 $, $ \gamma_{l} = 5.60 $ and full curve for $ n = 6 $,
 $ \gamma_{l} = 2.32 $.} \label{Fig-tr}
\end{figure}

\begin{figure}[h]
\vspace{0.8cm}
\includegraphics[height=3.0in,width=4.0in]{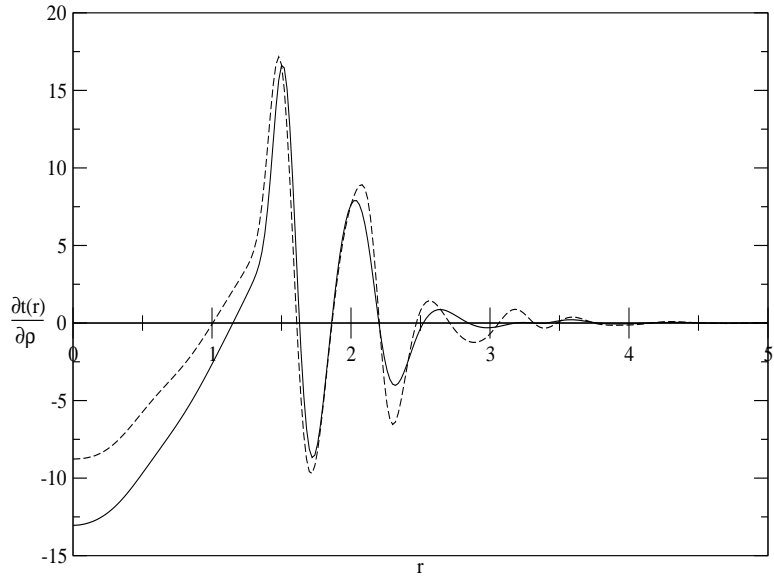}
\caption{Plot of $ \frac{\partial t(r)}{\partial \rho} $ vs $ r $
 for $ n = 6 $ at $ \gamma = 2.32 $ and $ n = 4 $ at $ \gamma = 5.60 $.
 Other notations are same as in Fig. 2. }  \label{Fig-dtr}
\end{figure}

\begin{figure}[h]
\vspace{0.8cm}
\includegraphics[height=3.0in,width=4.0in]{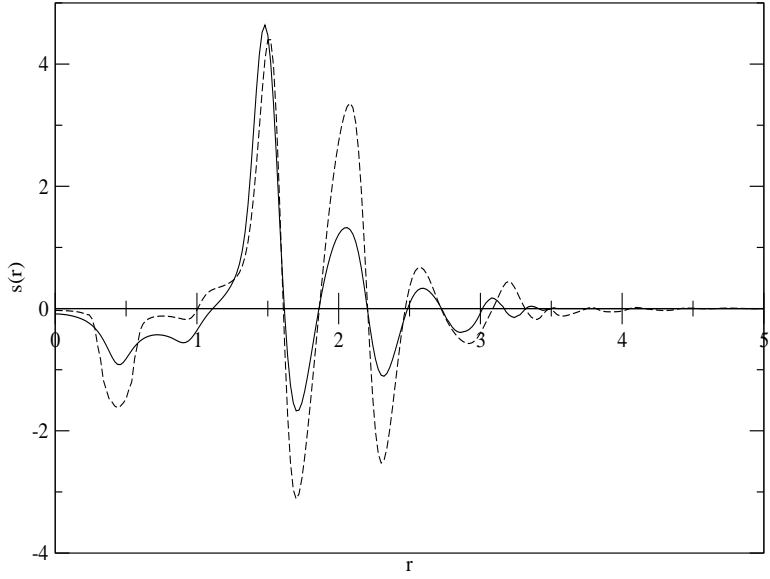}
\caption{Plot of $ s(r) $ vs $ r $ for $ n = 6 $ 
at $ \gamma = 2.32 $ and $ n = 4 $ at 
$ \gamma = 5.60 $. Other notations are same as in Fig. 2.} \label{Fig-sr}
\end{figure}

\begin{figure}[h]
\vspace{0.8cm}
\includegraphics[height=3.0in,width=4.0in]{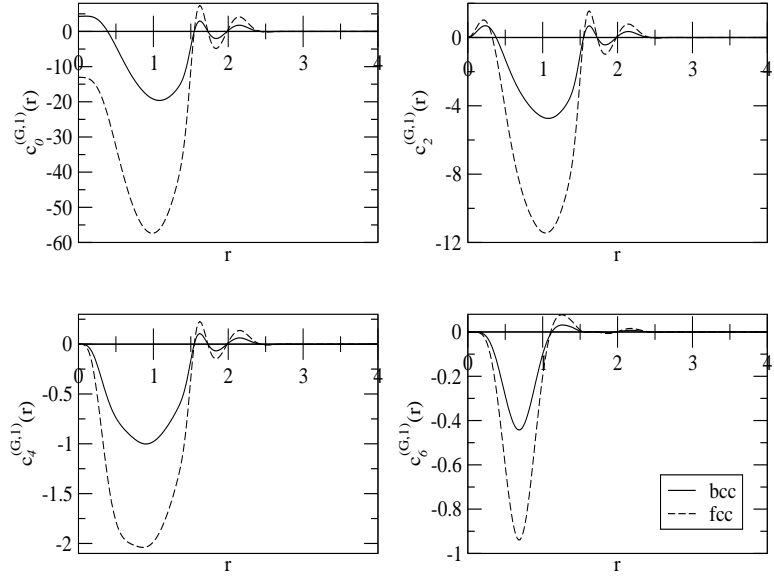}
\caption{Comparison of values of $ c_{l}^{(G,1)}(r) $ as a function
 of $r$ for a $ G $ vector of first set of fcc and bcc lattices for
 $ n = 6 $, $ \gamma_{s} = 2.32 $, $\alpha_{fcc} = 32 $ and 
 $\alpha_{bcc} = 18$. The distance $ r $ is in unit of 
 $ a_{0}=\left(\frac{3}{4 \pi \rho}\right)^{1/3}$ and 
 $ \mu = e^{-G^{2}/4\alpha} $. The dashed curve represents
 values of fcc  structure while full curve 
 of bcc structure. }  \label{Fig-cgr-3-1}
\end{figure}

\begin{figure}[h]
\vspace{0.8cm}
\includegraphics[height=3.0in,width=4.0in]{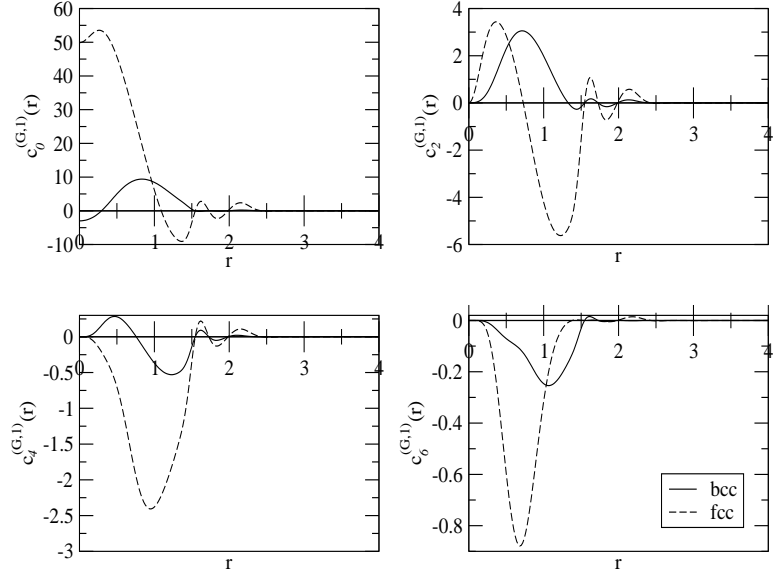}
\caption{Comparison of values of $ c_{l}^{(G,1)}(r) $ as a function
 of $r$ for a $ G $ vector of second set of fcc (dashed curve)
  and bcc (full curve) lattices for
 $ n = 6 $, $ \gamma_{s} = 2.32 $, $\alpha_{fcc} = 32 $ and 
 $\alpha_{bcc} = 18$. Other notations are same as in Fig. 5.}
 \label{Fig-cgr-3-2}
\end{figure}

\begin{figure}[h]
\vspace{0.8cm}
\includegraphics[height=3.0in,width=4.0in]{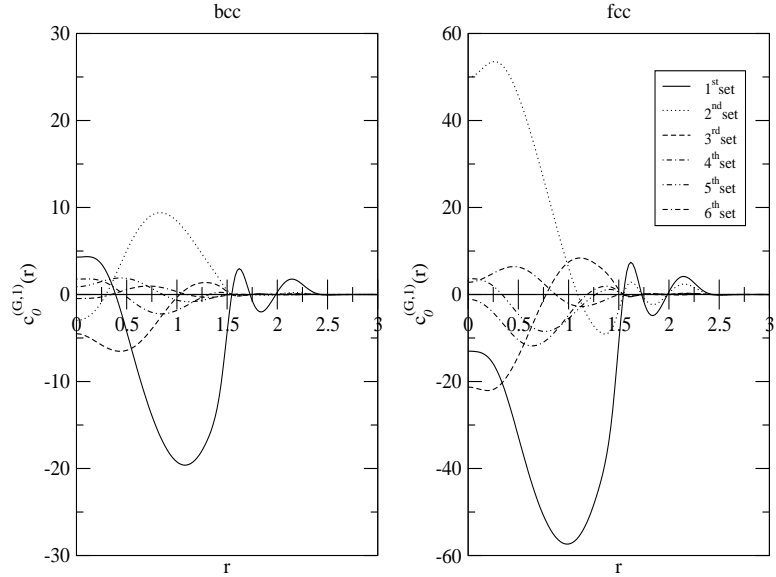}
\caption{Comparison of values of $ c_{0}^{(G,1)}(r) $ as a function
 of $r$ for a $ G $ vector of the first six sets of fcc and bcc lattices.
 The distance $r$ is in unit of 
 $ a_{0}=\left(\frac{3}{4 \pi \rho}\right)^{1/3}$.} \label{Fig-cgr-3-comp}
\end{figure}

\begin{figure}[h]
\vspace{0.8cm}
\includegraphics[height=2.5in,width=4.0in]{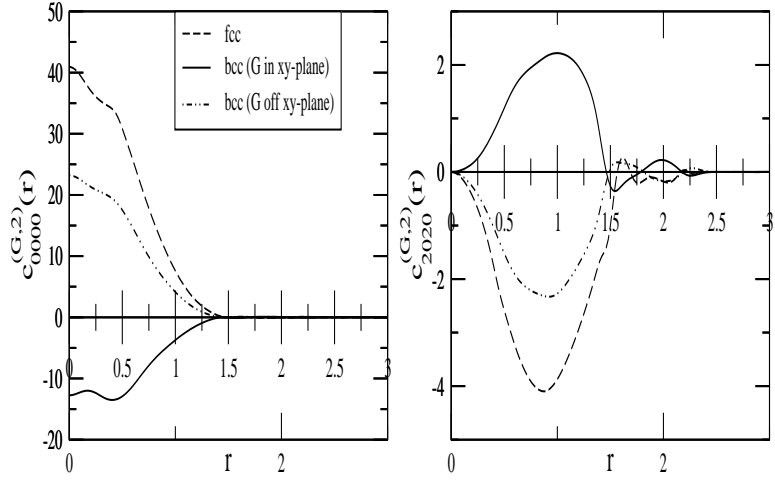}
\caption{Comparison of values of $ c_{lml'm'}^{(G,2)}(r) $ as a function
 of $r$ for a $ G $ vector of the first set of fcc and bcc lattices for
 $ n = 6 $ at $ \gamma_{s} = 2.32 $, $\alpha_{fcc} = 32 $ and 
 $\alpha_{bcc} = 18$. 
 The distance $r$ is in unit of $ a_{0}=\left(\frac{3}{4 \pi \rho}\right)^{1/3}$.
 There is two sets of values for bcc lattice; one for $ \vG $ vectors
 lying in x-y plane and the other for the rest of $ \vG $ vectors of the first set.
 There is only one set of values for fcc lattice.} 
 \label{Fig-cgr-4-1}
\end{figure}

\begin{figure}[h]
\vspace{0.8cm}
\includegraphics[height=2.5in,width=4.0in]{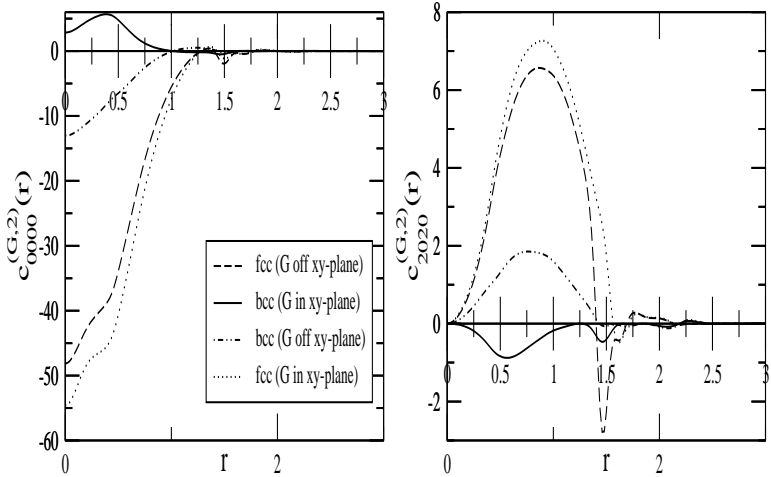}
\caption{Comparison of values of $ c_{lml'm'}^{(G,2)}(r) $ as a function
 of $r$ for a $ G $ vector of the second set of fcc and bcc lattices. 
 Other notations are same as in Fig. 8, except that there is now two sets
  of values shown by dashed and dotted curves for fcc lattice (see text)
  } \label{Fig-cgr-4-2}
\end{figure}

\begin{figure}[h]
\vspace{0.8cm}
\includegraphics[height=3.0in,width=4.0in]{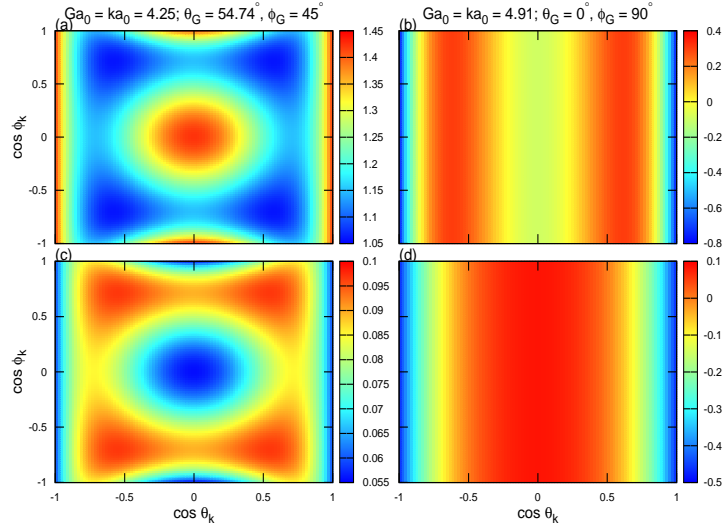}
\caption{Comparison of values (shown using color codes given
on right hand side of each figure) of 
 $ c^{(G,1)}(k,\theta_{k},\phi_{k}) $ and 
 $ c^{(G,2)}(k,\theta_{k},\phi_{k}) $ as a function of $ cos \theta_{k} $
 (plotted on x-axis) and $ cos \phi_{k} $ (plotted on y-axis) 
 for $G_{1}a_{0} = ka_{0} = 4.25$, $\theta_{G_{1}} = {54.7\degree}$,
 $\phi_{G_{1}} = {45\degree}$ (a and c) and $G_{2}a_{0} =
  ka_{0} = 4.91$, $\theta_{G_{2}} = {0\degree}$,
 $\phi_{G_{2}} = {90\degree}$ (b and d) for a fcc lattice 
 for $ n = 6 $, $ \gamma_{s} = 2.32 $, $ \alpha_{fcc} = 32 $.}
 \label{Fig-fcgkang}
\end{figure}

\begin{figure}[h]
\vspace{0.8cm}
\includegraphics[height=3.0in,width=4.0in]{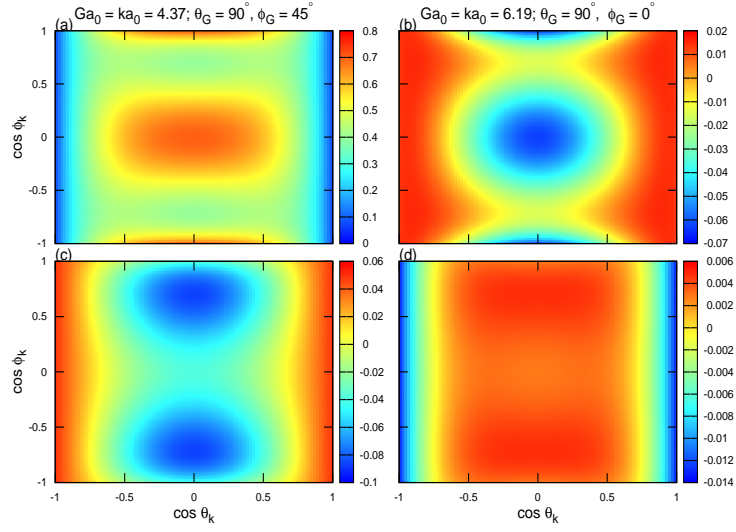}
\caption{Comparison of values (given in a color code) of 
 $ c^{(G,1)}(k,\theta_{k},\phi_{k}) $ and 
 $ c^{(G,2)}(k,\theta_{k},\phi_{k}) $ as a function of
 $ cos \theta_{k} $ and $ cos \phi_{k} $ for $G_{1}a_{0} =
 ka_{0} = 4.37$, $\theta_{G_{1}} = {90\degree}$,
 $\phi_{G_{1}} = {45\degree}$ (a and c) and $G_{2}a_{0} =
  ka_{0} = 6.19$, $\theta_{G_{2}} = {90\degree}$,
 $\phi_{G_{2}} = {0\degree}$ (b and d) for a bcc lattice 
 for $ n = 6 $, $ \gamma_{s} = 2.32 $, $ \alpha_{bcc} = 18 $.
 Other notation are same as in Fig. \ref{Fig-fcgkang}} \label{Fig-bcgkang}
\end{figure}

\begin{figure}[h]
\vspace{0.8cm}
\includegraphics[height=3.5in,width=5.0in]{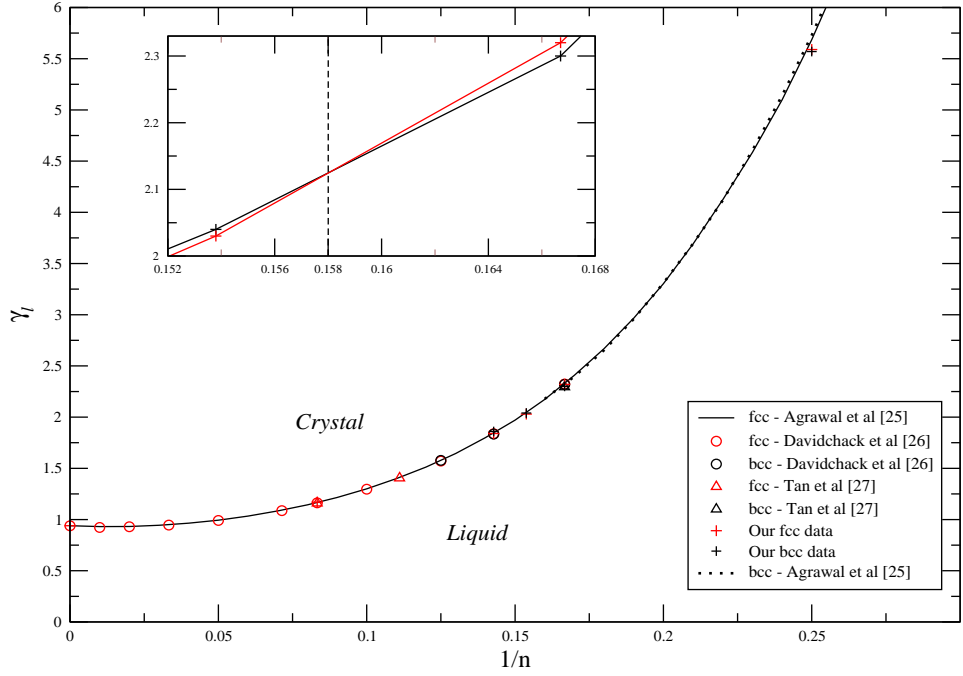}
\caption{Comparison of equilibrium phase diagram of $ \dfrac{1}{n} $
vs  $ \gamma_{l} $ found from simulation results and from our theory.
In inset the fluid-fcc and fluid-bcc transition lines are plated at a
magnified scale and the fluid-bcc-fcc triple point is found at 
$ \dfrac{1}{n} = 0.158 $. } \label{Fig-TP}
\end{figure}


\begin{figure}[h]
\vspace{0.8cm}
\includegraphics[height=3.5in,width=5.0in]{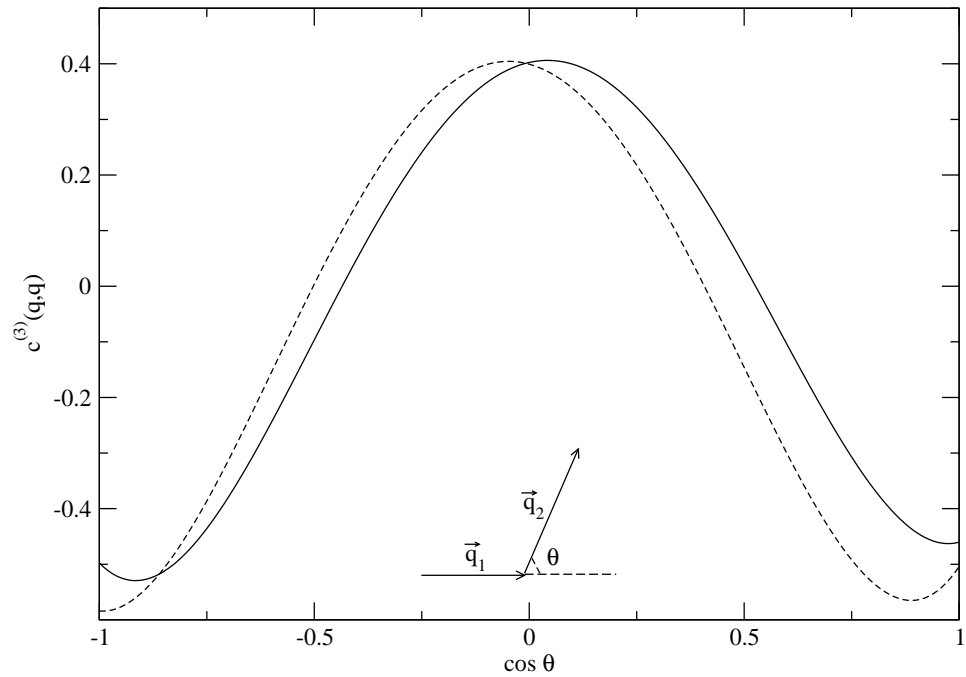}
\caption{Values of $ \hat{c}_{3}^{(0)}(q, q, q) $ as a function
of $ cos \theta $ (geometry is shown schematically in the figure)
for $ q_{1}a_{0} =  q_{2}a_{0} = 4.3$ and for potentials $ n = 12, 
\gamma_{l} = 1.17 $ (dashed curve) and $ n = 6, \gamma_{l} = 2.30 $
 (full curve)} \label{Fig-3bodycorr-1}
\end{figure}

\begin{figure}[h]
\vspace{0.8cm}
\includegraphics[height=3.5in,width=5.0in]{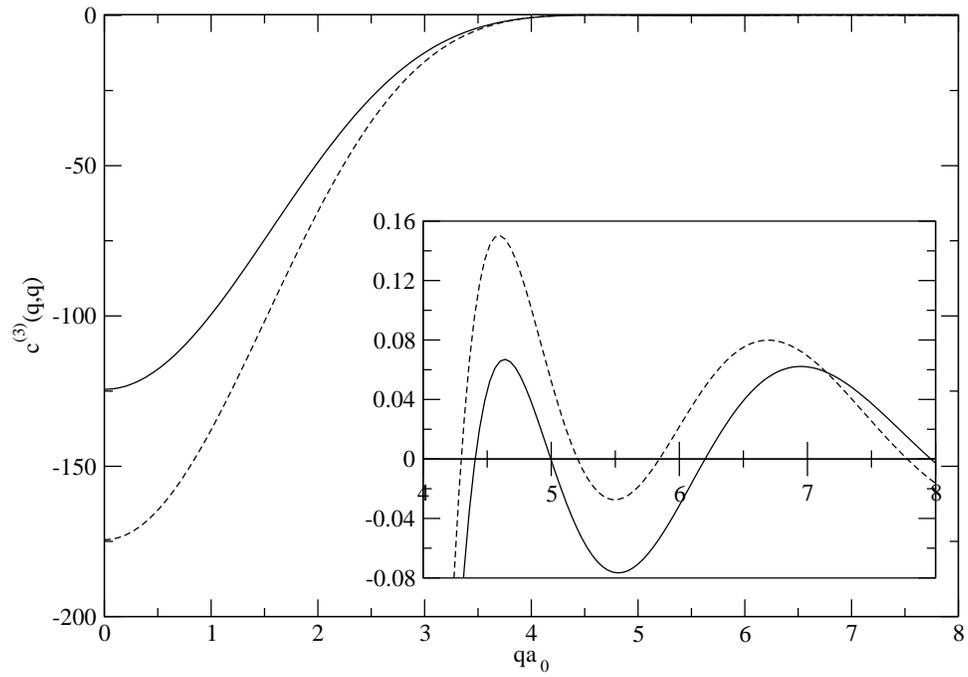}
\caption{Values of $ \hat{c}_{3}^{(0)}(q, q, q) $ vs $ qa_{0} $
(equilateral triangles). The dashed curve represents the values for 
 $ n = 12, \gamma_{l} = 1.17$ and full curve for $ n = 6, 
 \gamma_{l} = 2.30$. Inset shows values for $qa_{0} \geq 4.0$
 on magnified scale.} \label{Fig-3bodycorr-2}
\end{figure}

\begin{figure}[h]
\vspace{0.8cm}
\includegraphics[height=3.5in,width=5.0in]{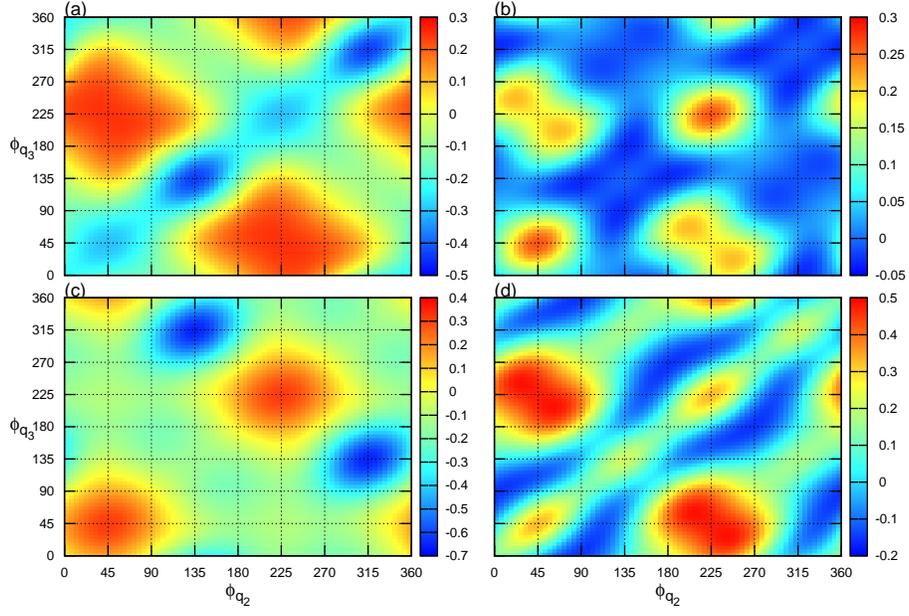}
\caption{ Values of $ \hat{c}_{4}^{(0)}(\vq_{1}, \vq_{2}, \vq_{3}) $
 (shown using a color code shown on right hand side of each figure)
 as a function of $ \phi_{q_{2}} $ and $ \phi_{q_{3}} $ for $ q_{1}
  = q_{2} = q_{max} $ with $ q_{max}a_{0} = 4.3 $, 
  $ \theta_{q_{1}} = 0^{\circ}$, $ \phi_{q_{1}} = 0^{\circ}$: In
 (a) $ \theta_{q_{2}} = 45^{\circ}$, $ \theta_{q_{3}} = 45^{\circ}$,
 (b) $ \theta_{q_{2}} = 45^{\circ}$, $ \theta_{q_{3}} = 90^{\circ}$,
 (c) $ \theta_{q_{2}} = 90^{\circ}$, $ \theta_{q_{3}} = 45^{\circ}$, and
 (d) $ \theta_{q_{2}} = 90^{\circ}$, $ \theta_{q_{3}} = 90^{\circ}$}
  \label{Fig-4bodycorr-1}
\end{figure}

\begin{figure}[h]
\vspace{0.8cm}
\includegraphics[height=3.5in,width=5.0in]{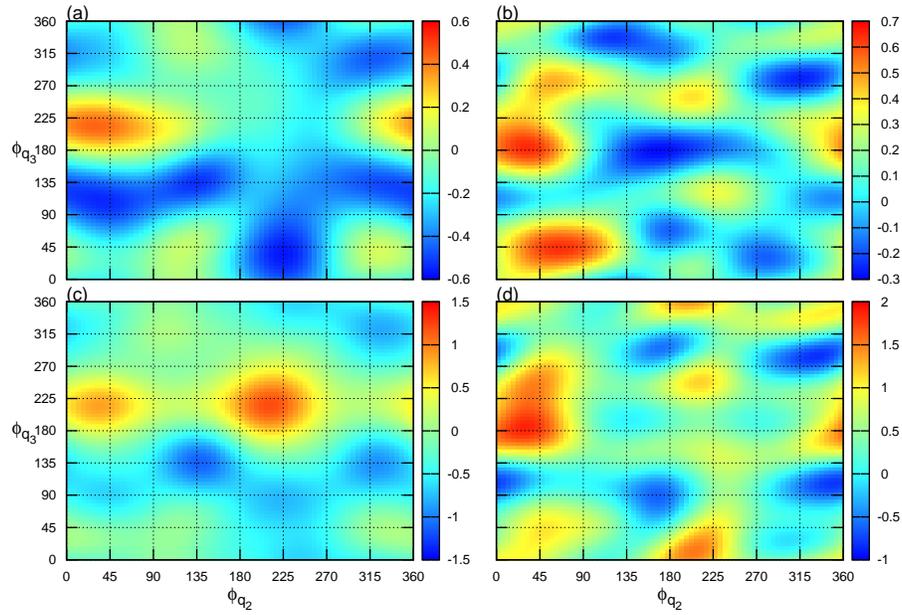}
\caption{Same as in Fig. \ref{Fig-4bodycorr-1} except $ \theta_{q_{1}} =
 90^{\circ} $ and $ \phi_{q_{1}} = 0^{\circ} $} \label{Fig-4bodycorr-2}
\end{figure}

\end{document}